\documentclass[aps,twocolumn,groupaddress,showpacs,color]{revtex4-1}
\usepackage{graphicx}
\usepackage{amssymb}
\usepackage{amsmath}
\usepackage{amsfonts}
\newcommand{\model}{Eqs.~(\ref{spp1}, \ref{spp2}) }

\begin{document}

\title{Clustering and heterogeneous dynamics in 
a kinetic Monte-Carlo model of self-propelled hard disks}

\author{Demian Levis}

\affiliation{Laboratoire Charles Coulomb, UMR 5221 CNRS and 
Universit\'e Montpellier 2, Montpellier, France}

\author{Ludovic Berthier}

\affiliation{Laboratoire Charles Coulomb, UMR 5221 CNRS and 
Universit\'e Montpellier 2, Montpellier, France}

\date{\today}

\begin{abstract}
We introduce a kinetic Monte-Carlo model 
for self-propelled hard disks to capture with minimal ingredients 
the interplay between thermal fluctuations, excluded volume and self-propulsion
in large assemblies of active particles.
We analyze in detail the resulting (density, self-propulsion) nonequilibrium 
phase diagram over a broad range of parameters. 
We find that purely repulsive hard disks spontaneously 
aggregate into fractal clusters as self-propulsion is increased, 
and rationalize the evolution of the average cluster size
by developing a kinetic model of reversible aggregation.
As density is increased, the nonequilibrium clusters percolate to 
form a ramified structure reminiscent of a physical gel. 
We show that the addition of a finite amount of noise is needed 
to trigger a nonequilibrium phase separation, showing that 
demixing in active Brownian 
particles results from a delicate balance between noise, interparticle 
interactions and self-propulsion. We show that self-propulsion has a 
profound influence on the dynamics of the active fluid. 
We find that the diffusion constant 
has a nonmonotonic behaviour as self-propulsion is increased
at finite density and that activity produces strong deviations 
from Fickian diffusion that persist over large time scales and length scales,
suggesting that systems of active particles generically behave as 
dynamically heterogeneous systems.
\end{abstract}



\maketitle

\section{Introduction}

In many active materials, the relevant entities 
(molecules, cells, animals...) are self-propelled objects 
which can borrow energy 
from their environment to produce their own motion using some internal 
mechanism~\cite{Ramaswamy2010,Vicsek2012,Marchetti2012a}. At the 
fundamental level, the local energy input required by self-propulsion 
breaks detailed balance, which automatically drives 
the system far from thermodynamic equilibrium.
Self-propulsion is therefore directly responsible 
for the emergence of a number of collective phenomena that are not 
observed in the absence of activity, such as, for instance, the existence of an orientationally ordered state in two spatial 
dimensions~\cite{Toner2005,Vicsek2012,Marchetti2012a,Ramaswamy2010}, 
the formation of coherently moving patterns, aggregation, clustering, motility-induced phase separation and giant number 
fluctuations~\cite{Toner2005,Ramaswamy2010,Vicsek2012,Marchetti2012a,Peruani2006,Ginelli2010,Tailleur2008,CatesRev,Filly2012,Bricard2013}. 

Self-propelled `particle' systems are manifold in biology, 
with examples ranging from animal groups to bacterial colonies 
or molecular motors in the cytoskeleton~\cite{Vicsek2012}. 
Recently, artificial or `abiotic' physical systems of self-propelled particles have been realised in the laboratory using, for instance, granular materials~\cite{Narayan2007,Deseigne2010} or colloidal particles with specific surface treatments~\cite{Paxton2004,Granick2006simple,Granick2006clusters,Volpe2011,Theurkauff2012,Palacci2013,Bechinger2013}. 
These experimental developments  offer an ideal playground to investigate general features of self-propelled particle systems since they are simpler to control and more versatile than their biological counterparts.
Because the physics of granular and colloidal particles has been analyzed in great detail, self-propulsion can be seen as a new physical ingredient whose influence needs to be studied. In this work, our central objective is to understand how deviations from detailed balance originating from persistent self-propulsion 
affects the structure and dynamics of simple fluids.

Over the last few years, the field of active matter has been the subject of a large number of studies, striking interest from a broad 
community of researchers stemming from different 
fields~\cite{Vicsek2012,Marchetti2012a,Ramaswamy2010}.  
Previous studies have revealed the importance of the shape and polarity of self-propelled particles for their collective behaviour. Agent-based models of polar particles with an alignment interaction undergo a phase transition towards a polar ordered state, resulting in giant number fluctuations~\cite{Toner2005,Zhang2010,Ginelli2010}. Elongated apolar self-propelled particles such as granular rods or myxobacteria with excluded volume interactions order in a nematic state with no polar collective motion. Moreover, nematic particles self-organise into coherently moving patterns~\cite{Chate2006,Peruani2006,Narayan2007}. Closer to simple liquids and to our own approach, 
recent studies have shown that systems made of apolar isotropic particles with no orientational order may exhibit a motility-induced 
phase separation~\cite{Tailleur2008,Filly2012,Redner2013}, 
which has motivated a large number of 
studies~\cite{Thompson2011,Bialke2013EPL,Cates2013EPL,Marenduzzo2013,Stenhammar2014,Wittkowski2013}. 

The recent development of simple experimental systems of self-propelled granular and colloidal particles motivates the study of simple model systems where the influence of self-propulsion on the equilibrium structure and dynamics of simple fluids can be understood in detail.
The hard sphere model is perhaps 
the simplest and most studied model to study the physics of 
simple liquids~\cite{HansenBook},
where physics stems from the competition between hard core repulsion and 
thermal fluctuations. 
Hard spheres represent also an excellent model to understand the 
physics of colloidal and granular particles. Therefore, we 
decided to develop a simple model of self-propelled hard particles, 
specializing ourselves to two dimensions where the majority of 
experiments on active particles are performed,
but the model is easily generalized to treat higher dimensions. 

So far, the statistical mechanics of 
active particles has been studied numerically using mainly molecular dynamics 
simulations, where self-propulsion is introduced in Langevin
descriptions of active Brownian dynamics by considering 
specific couplings between translational and orientational 
motion~\cite{Loi2008,Bialke2012,CatesRev,Filly2012,Redner2013}. 
In these descriptions, more complicated phenomena induced
by additional coupling to hydrodynamic fluctuations are neglected, 
but they still contain an appreciable number of control parameters
that must be simultaneously adjusted. 
Here, we wish to develop a minimal strategy to capture at the
most basic level how self-propulsion affects the structure and
dynamics of simple fluids, extending the approach successfully used 
to study simple fluids at thermal equilibrium, namely, 
kinetic Monte-Carlo simulations~\cite{NewmanBook,Landau2009Book}. 
To our knowledge, Monte-Carlo 
simulations of active particle systems have not been reported so far.
We will therefore carefully explain our modelling, justify 
its relevance for the field of active matter
and compare it to alternative models in some detail.  

The principal aim of our work is therefore to develop a 
`minimal' numerical model with as few control parameters 
as possible. To this end, we introduce
a kinetic Monte-Carlo model of self-propelled hard disk, where activity 
is controlled by {\it a single control parameter}, which 
reduces, in the dilute limit,
to the persistence time $\tau$ of a persistent random walk 
motion. Equivalently, this control parameter can be seen as 
a `rotational' P\'eclet number~\cite{Gompperpeclet,Starkpeclet}. 
Because thermal fluctuations 
only affect rotational degrees of freedom, 
the `translational' P\'eclet number is not a convenient 
control parameter in our model~\cite{Filly2012}.
The second control parameter of the model
is the packing fraction $\phi$ of the hard disk system, 
which is the unique one in equilibrium conditions.
Although our model is presumably too simple to describe the details of 
any specific experimental realization, we shall demonstrate 
that it can still capture some essential features of 
the interplay between thermal fluctuations, 
excluded volume, and self-propulsion.

\begin{figure}
\includegraphics[scale=0.42,angle=0]{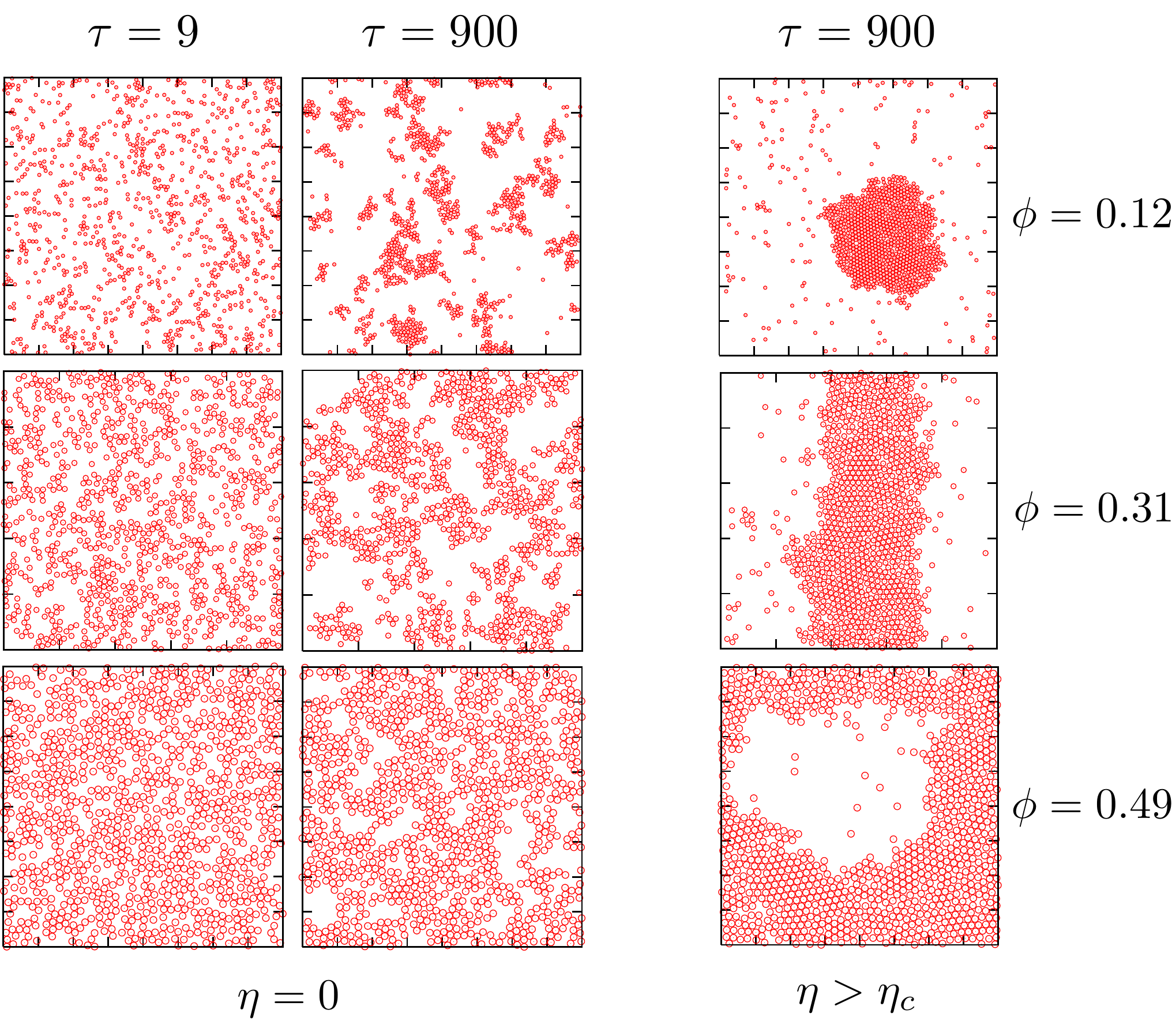}
\caption{Representative
snapshots of a system composed of $N=10^3$ self-propelled hard disks 
at different packing fractions $\phi$. Left and central columns 
correspond to two different values of the persistence time $\tau$, 
showing a fluid phase at low $\phi$ and low $\tau$, the emergence
of clusters as $\tau$ is increased, and a gel-like structure 
at large $\phi$ and large $\tau$. 
The right column shows phase separated systems obtained upon the addition
of translational noise of finite amplitude $\eta$ across a broad 
range of densities.}
\label{fig:snapshotsT0}
\end{figure}

The present work presents a detailed study of the 
$(\phi, \tau)$ phase diagram of the model, which reduces
to the known equilibrium hard disk fluid model 
in the limit $\tau \to 0$. The glassy active dynamics
of the model at large density
is studied elsewhere~\cite{Berthier2013}, 
and we concentrate in this work 
on the regime $0 \leq \phi \leq 0.60$.  
Our main results are summarized in Fig.~\ref{fig:snapshotsT0},
which displays the emergence, at any finite density, 
of complex nonequilibrium structures in the active fluid, taking the form 
of finite size clusters at moderate density that percolate
to form a gel-like structure at larger density. These structural 
changes are accompanied by profound changes in the dynamics 
of the system, which we also analyse in detail. Surprisingly, 
clustering in our model does not take the form of a macroscopic 
phase separation found in previous numerical models of active Brownian
particles~\cite{Tailleur2008,Filly2012,Redner2013,FilyMarchetti2013,Gompper2013}. 
We elucidate 
the physical origin of this qualitative difference by introducing 
an additional translational noise of strength $\eta$ in our model, and by
showing that phase separation can only occur in a restricted part of 
the extended $(\phi, \tau, \eta)$ phase diagram where $\eta >0$ 
(see the corresponding snapshots in Fig.~\ref{fig:snapshotsT0}).
This demonstrates that phase separation in fact results 
from a balance between noise, interparticle 
interactions and self-propulsion that is more delicate and 
perhaps less generic than previously thought. An interesting corollary 
is that clustering in our model can be interpreted as 
the result of a kinetically arrested phase separation, which adds a novel
possible mechanism to the ongoing debate about the origin of clustering 
in self-propelled particle systems~\cite{Theurkauff2012,Filly2012,FrenkelValeriani2013,BallAPS}.  

The paper is organised as follows. 
In Sec.~\ref{model}, we define our kinetic 
Monte-Carlo approach to simulate the behaviour of self-propelled hard disks, 
present the control parameters, units, behaviour in the dilute
limit, and comparison with existing models.
In Sec.~\ref{structure} we establish the 
phase diagram of the model in steady state conditions, showing 
fluid, clustered and percolated phases whose boundaries
are carefully studied. We construct
and solve a kinetic model that reproduces the cluster size distribution 
obtained in the simulations and gives useful insights into the 
aggregation mechanism.
In Sec.~\ref{sec:phsep} we show that a finite amount of 
translational noise triggers a macroscopic phase separation
over a broad range of densities.
In Sec.~\ref{dynamics} we analyse the dynamics of the model 
in steady state, and show that the relaxation of the system is 
highly heterogeneous. 
In Sec.~\ref{discussion} we summarize and discuss our results. 

\section{Kinetic Monte-Carlo model for self-propelled 
hard disks} 

\label{model}

\subsection{Hard disk model: Volume fraction $\phi$}

We work with a bidimensional,
monodisperse assembly of $N$ hard disks of diameter $\sigma$ 
enclosed in a square box of linear size $L$, using 
periodic boundary conditions. The hard core interaction implies 
that no overlap between disk is allowed, whereas configurations with 
no overlap all have the same energy, which can be chosen to be zero, 
for convenience. Therefore, by contrast with most 
works dealing with active Brownian particles, the particle 
softness is not an independent control parameter, and density 
and temperature cannot be independently adjusted because we use 
an infinitely hard core repulsion.
The system is thus uniquely characterized by the packing (or more 
precisely, area) fraction, 
\begin{equation}
\phi = \frac{\pi N \sigma^2}{4 L^2}. 
\end{equation}

\subsection{Kinetic Monte-Carlo in equilibrium}

\label{equilibrium}

Monte-Carlo simulations are traditionally viewed  as 
an efficient way to sample the configurational phase space
with a given probability distribution, usually
chosen as the Boltzmann distribution~\cite{NewmanBook}. However, 
it is well known that Monte-Carlo simulations can also 
be used to analyse the kinetics of statistical models, including 
off-lattice complex fluids~\cite{Landau2009Book}. 

For hard disks, the equilibrium Monte-Carlo approach 
is conceptually very simple, and proceeds as follows. 
The positions of the particles are updated sequentially, 
and dynamics results from the repetition of 
the following elementary Monte-Carlo trial step. 

\begin{itemize}

\item 
At time $t$, a particle is chosen at random, say 
particle $i$. 

\item
A random displacement ${\vec \delta_i}(t)$ is drawn
from a chosen distribution. We shall use the following  
convention: 
\begin{equation}
{\vec \delta_i}(t) = \delta_0 {\vec \xi_i}(t),
\label{eqmc1}
\end{equation}
where ${\vec \xi_i}(t)$ is a random vector with components 
independently drawn from a flat distribution in the interval $[ -1, 1]$,
so that typical displacements have a typical amplitude $\approx 
\delta_0$.

\item
The move is accepted if it creates no overlap between 
the disks.
This trial Monte-Carlo step can be written as
\begin{equation}
{\vec r_i}(t+1) = {\vec r_i}(t) + {\vec \delta_i}(t) p_{\rm acc}(t),
\label{eqmc2}
\end{equation}
where the acceptance probability $p_{\rm acc}(t)$ is unity 
if no overlap is created, zero otherwise. 

\end{itemize}

Conventionally, a Monte-Carlo time step $\tau_{\rm MC}$ 
is defined as the succession of $N$ such elementary 
moves, so that the dynamical behaviour expressed in units 
of $\tau_{\rm MC}$ does not depend 
on the number of particles in the limit of large $N$~\cite{Landau2009Book}.  

Another interesting limit is when 
the elementary step size $\delta_0$ becomes very small. 
If this limit is considered, it is more convenient to use 
a different time unit, 
$\tau_{\rm MC}' = \sqrt{\delta_0} \tau_{\rm MC}$, 
such that the dynamics becomes independent of $\delta_0$ 
as $\delta_0 \to 0$. In this limit, the above 
Monte-Carlo dynamics becomes equivalent to 
the following Brownian dynamics, 
\begin{equation}
\frac{\partial}{\partial t} 
{\vec r_i}(t) = \sum_j {\vec f}_{ij}(t) + {\vec \zeta_i}(t), 
\end{equation}
where ${\vec f}_{ij}$ is the interparticle force, and 
${\vec \zeta_i}$ a Gaussian random noise satisfying 
the fluctuation-dissipation theorem, 
$\langle \zeta_{i,\alpha}(t) \zeta_{j,\beta}(t') \rangle = 2 \delta_{\alpha \beta}
\delta_{ij} \delta(t-t')$.

\subsection{Introduction of self-propulsion: Persistence time $\tau$} 

To introduce self-propulsion in the hard disk model, 
we must produce time correlations in the elementary
particle displacements.  
A simple way to do this is to introduce correlations
between successive displacements of the particles. 
To this end, we generalize Eqs.~(\ref{eqmc1}, \ref{eqmc2}) to  
\begin{eqnarray}
{\vec \delta_i}(t) & = & {\vec \delta_i}(t-1) + \delta_1 {\vec \xi_i}(t), 
\label{spp1}
\\
{\vec r_i}(t+1) & =  & {\vec r_i}(t) + {\vec \delta_i}(t) p_{\rm acc}(t),
\label{spp2}
\end{eqnarray}
where we enforce the condition that $|\delta_{i,\alpha}(t)| \leq \delta_0$, 
just as in the original equilibrium dynamics in Eq.~(\ref{eqmc1}).   

The physics of the kinetic Monte-Carlo model in Eqs.~(\ref{spp1}, \ref{spp2})
is transparent, as it differs from the equilibrium Monte-Carlo 
dynamics only by the fact that the random displacement ${\vec \delta_i}(t)$ 
now performs a simple random walk of average jump 
length $\delta_1$ in a square of linear 
size $\delta_0$ (with reflective boundary conditions). 

The interesting regime is when $\delta_1 < \delta_0$, 
where the orientation of the elementary displacement then decorrelates
slowly in about $(\delta_0 / \delta_1)^2$ trial moves (this scaling 
stems from the fact that ${\vec \delta_i}$ must diffuse
a distance $\delta_0$ using independent steps of size $\delta_1$).    
This implies that the elementary trial moves develop 
time correlations, and decorrelate after a 
typical persistent time $\tau$,
\begin{equation}
\tau = \left( \frac{\delta_0}{\delta_1} \right)^2 \tau_{\rm MC},
\label{tau}
\end{equation}
which defines the second control parameter of the model. 
This timescale has sometimes been named {\it rotational P\'eclet 
number}~\cite{Gompperpeclet,Starkpeclet}, 
as it quantifies the efficiency of the fluctuations that 
are responsible for the rotation of the displacement, as 
modelled by the term $\delta_1 {\vec \xi_i}$ in Eq.~(\ref{spp1}).
Note that it is not convenient to define a 
P\'eclet number 
based on translational degrees of freedom from Eq.~(\ref{spp2}),
a situation encountered in previous models~\cite{Filly2012}.

The key feature of the kinetic model defined by Eqs.~(\ref{spp1}, \ref{spp2})
is that whereas the model is Markovian in the enlarged space of 
positions and displacements, it becomes non-Markovian 
and time irreversible in the subspace of the positions, where detailed balance
is therefore broken. As a result, the model becomes purely `dynamical'
in the sense that 
the configurational space of the hard disk system is not sampled 
according to the Boltzmann distribution anymore. In short, 
self-propulsion pushes the system far from equilibrium,
which is only recovered in the limit $\tau \to 0$ where 
equilibrium sampling as in Eq.~(\ref{eqmc1}) is obtained. 
By slowly increasing $\tau$ we can then observe how the system
departs from a well-known equilibrium situation. Such a smooth connection
to equilibrium is not always possible in earlier models of active 
particles~\cite{Filly2012,FilyMarchetti2013}.  

In practive we have studied system sizes 
in the range $N = 10^3 -10^4$, to test against possible finite size 
effects which will be discussed whenever they are relevant.
We have changed the persistence 
time $\tau$ in Eq.~(\ref{tau}) using a fixed length
scale $\delta_0 = 0.1$ controlling the maximal size of the elementary moves, 
and by varying $\delta_1$ in Eq.~(\ref{spp1}). We have explored 
a regime $\delta_1 = 0.001 - 0.1$, thereby covering 
a range of persistent times of 4 orders of magnitude, 
$\tau / \tau_{\rm MC} = 1 - 10^4$.  

From now on, we use the particle diameter $\sigma$ as the unit length scale, 
and the Monte-Carlo time step $\tau_{\rm MC}$ as the unit time scale. 
 
\subsection{Dilute limit: persistent random walk}

\label{subdilute}

The physics of the dilute limit is straightforward. 
The regime where $\tau \ll 1$ is trivial since 
particles then perform a simple random walk, with a jump size 
controlled by $\delta_0$, in which case, the 
mean-squared displacement defined as
\begin{equation}
\Delta^2(t) = \langle [ {\vec r_i}(t)- {\vec r}_i(0) ]^2\rangle
\label{msd}
\end{equation}
increases as $\Delta^2(t) = D_0 t$, with $D_0 \approx \delta_0^2$.  

Let us consider the more interesting
regime where $\tau \gg 1$. Two time regimes 
have then to be considered. When $t \ll \tau$, displacements
are nearly persistent and the mean-squared displacement 
increases ballistically as 
$\Delta^2(t) \approx (\delta_0 t)^2.$  
On the other hand, the displacements become uncorrelated 
for times $t \gg \tau$, and motion becomes diffusive,
$\Delta^2 \approx D t$, with 
\begin{equation}
D \approx D_0 \tau.
\end{equation}  
This shows that, as expected in the dilute limit
where particles do not interact, the self-propulsion mechanism enhances 
self-diffusion, simply because ballistic motion
is more efficient than diffusion.  

\begin{figure}
\centering
\includegraphics[scale=0.47,angle=0]{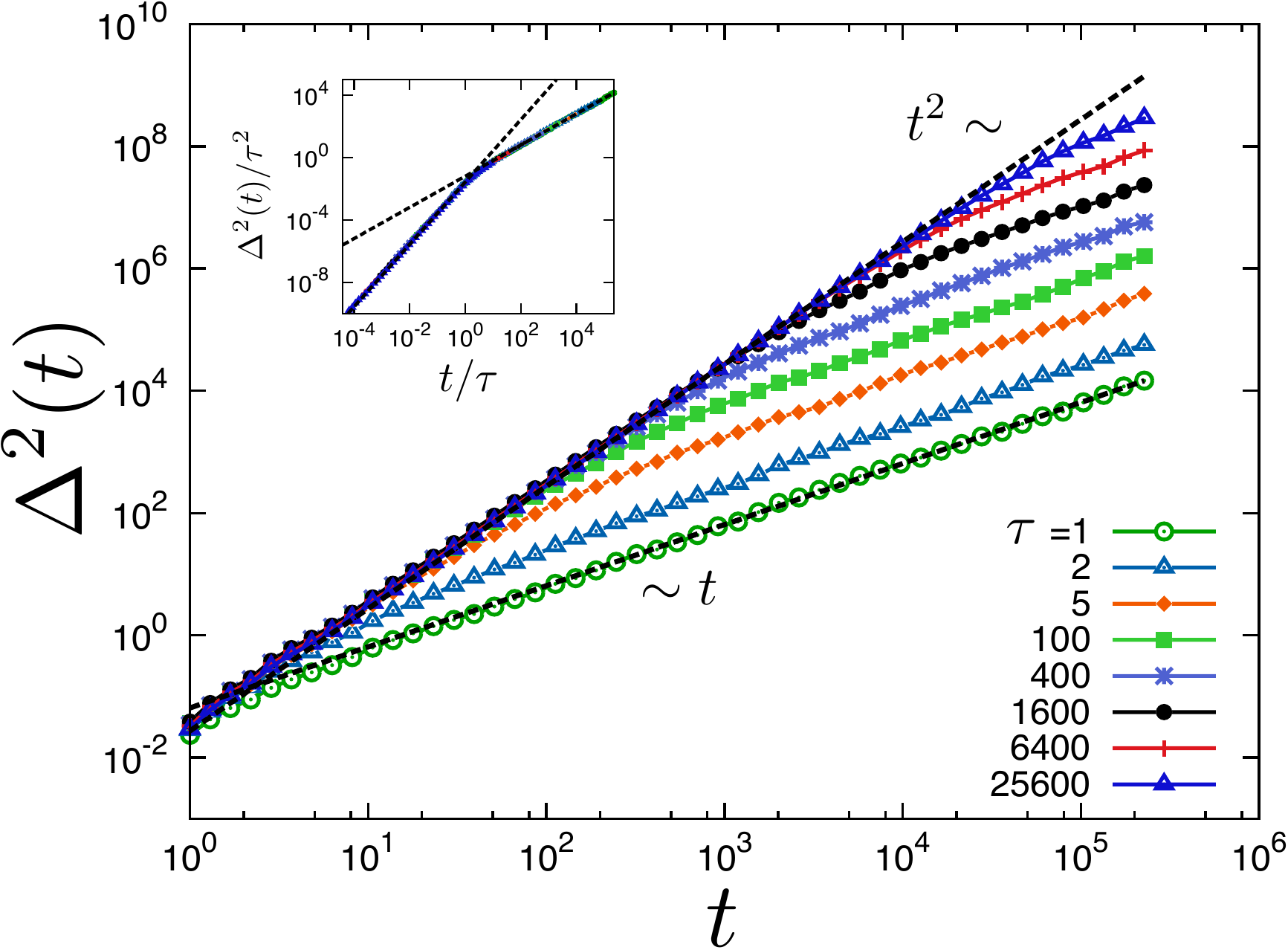}
\caption{Time dependence of the mean-squared displacement, Eq.~(\ref{msd}), 
in the dilute limit, $\phi \to 0$, for different persistent times $\tau$,
showing a crossover from ballistic to diffusive motion, 
with a diffusion constant that increases with $\tau$.  
Inset: The same data are collapsed using the scaling form in 
Eq.~(\ref{eq:MSD_SC}) with ballistic and diffusive asymptotics indicated 
with dashed lines.}
\label{fig:MSDPF0}
\end{figure}

In Fig.~\ref{fig:MSDPF0}, we present numerical data obtained from 
simulating the kinetic Monte-Carlo model Eqs.~(\ref{spp1}, \ref{spp2}) 
in the dilute limit, $\phi \to 0$. These data confirm 
the above description, with the observation of ballistic motion
crossing over to diffusive behaviour at a time scale controlled 
by $\tau$, therefore reducing to simple diffusion in the 
equilibrium limit $\tau \to 0$.  Because this crossover
between ballistic and diffusive motion is uniquely controlled 
by the persistence time $\tau$, the 
mean-squared displacement takes the following scaling 
form,  
\begin{equation}
\Delta^2(t)= \tau^2F(t/\tau), \label{eq:MSD_SC}
\end{equation} 
where $F(x) \sim x^2$ for $ x \ll 1$ and 
$F(x)\sim  x$ for $x \gg 1$. This scaling form 
suggests that data can be collapsed when $\Delta^2 / \tau^2$
is represented as a function of the rescaled time $t /\tau$, 
as demonstrated in the inset of Fig.~\ref{fig:MSDPF0}. 
Physically, Eq.~(\ref{eq:MSD_SC}) also implies that 
in the dilute limit, it is equivalent to vary the persistence time 
or the velocity characterizing the ballistic motion at short time. 

\subsection{Adding fluctuations: Translational noise $\eta$}

\label{addingeta}

By definition of the kinetic model in Eqs.~(\ref{spp1}, \ref{spp2}),  
elementary particle displacements ${\vec r_i}(t+1) - {\vec r_i}(t)$
are fully slaved to the orientation of the vector 
${\vec \delta_i}(t)$, whose time evolution is not affected 
by the particle motion. Thus, in our model, particles have only 
two choices: either move in the direction imposed by ${\vec \delta_i}(t)$
(when $p_{acc}(t)=1$), or stop (when $p_{acc}(t)=0$). 

The main body of the article is devoted to studying 
the behaviour of the model using these simple kinetic rules. 
However, to establish connections with other models proposed for 
active Brownian particles~\cite{Filly2012,Redner2013,Marenduzz2013}, 
it will be useful to introduce 
an additional translational noise in our model. 
There are many possible ways to do this, and we have checked that 
our results do not depend crucially on the specific choice we made. 

We perturb the self-propelled dynamics 
in Eqs.~(\ref{spp1}, \ref{spp2}) by performing additional
Monte-Carlo moves of small amplitude $\delta_\eta$ that have the same 
Markovian properties as in equilibrium dynamics. 
In practice, for each Monte-Carlo trial, we choose 
with probability 1/2 to use the self-propelled 
dynamics in Eqs.~(\ref{spp1}, \ref{spp2}), or to 
draw a random displacement $\delta_\eta {\vec \xi_i}(t)$ 
(see Eq.~(\ref{eqmc1})). This perturbation then allows particle 
displacements that are not slaved
to the direction of ${\vec \delta_i}(t)$, and 
therefore it introduces noise in our kinetic Monte-Carlo model. 

We quantify the strength of the noise by the quantity
$\eta = \delta_\eta / \delta_0$, which compares 
the relative size of the Monte-Carlo moves performed
without memory (of amplitude $\delta_\eta$) to the 
persistent moves (of amplitude $\delta_0$). 
Our original kinetic 
Monte-Carlo model is recovered as $\eta \to 0$, while 
for finite $\eta$, the model then lives in an
extended three-dimensional phase diagram $(\phi, \tau, \eta)$.  

Our primary motivation to introduce $\eta$ as an additional 
control parameter is to make contact with earlier numerical 
results, as detailed below.
In particular, we emphasize that $\tau$ and $\eta$ separately 
control the strength of the self-propulsion and that of the noise,
which will allow us to disentangle the relative role they might play 
in the physics of active Brownian particles. 
 
\subsection{Comparison with earlier models}

\label{comparisonmodel}

The kinetic rules defining our Monte-Carlo model in \model
fall in between two types of modelling that have been put forward
to understand the physics of active particle systems.

Because dynamics proceeds in discrete time steps, 
our model bears similarities with approaches initiated  
by Vicsek and coworkers, which analyse the dynamics of point 
particles evolving under the influence of aligning interactions 
and noise~\cite{Vicsek2012}. Although several distinct versions of 
such models have been studied 
in the literature, we are not aware of any Vicsek-like numerical 
model system incorporating physical ingredients comparable to ours.

A different class of models is obtained starting from 
Langevin equations governing the time evolution 
of the position and orientation of the self-propelled 
particles~\cite{Filly2012,Redner2013,Marenduzz2013}.
Because we can take the limit of small step sizes in the kinetic Monte-Carlo
approach, we expect our model to also bear strong similarities with   
Langevin models neglecting inertia, hydrodynamic interactions, 
and particle anisotropy, as studied for instance in 
Refs.~\cite{Redner2013,Filly2012,Bialke2012,FilyMarchetti2013,Ni2013}. 
It is therefore interesting to compare 
more precisely our model to this family of earlier studies. 

To proceed we must first discuss the small step size 
limit in our model, in analogy with the discussion of equilibrium 
dynamics in 
Sec.~\ref{equilibrium}. To analyse the limit $\delta_0 \to 0$, 
it is useful to introduce, again, a distinct time unit,
$\tau_{\rm MC}'' = \delta_0 \tau_{\rm MC}$. In these novel (double-primed)
units,
the persistence time becomes $\tau'' = \delta_0 \tau$,
persistent motion at short times $t'' \ll \tau''$ takes place
with a velocity $v''=1$, while the long-time diffusion 
coefficient controlling the regime $t'' \gg \tau''$
in the dilute limit becomes $D'' = \tau''$. 
The limit $\delta_0$ can then safely be taken, the physics 
remaining unaffected by the chosen $\delta_0$ value, 
provided the persistence time 
$\tau''$ is kept fixed. In practical terms, this implies 
taking the limits $\delta_0 \to 0$ and $\delta_1 \to 0$, while 
keeping the ratio $\delta_0^3/\delta_1^2 = 
\tau''$ constant. 

This reasoning applies as long as $\delta_0$ is small compared to any other 
length scale in the problem. We have checked that 
the numerical results presented in this paper are not 
affected by the specific choice $\delta_0 = 0.1$ we made. 
This choice is dictated by the usual trade-off between having a very small 
$\delta_0$ to get closer to the continuous time limit, 
and the large jumps needed 
to make simulations more efficient~\cite{Berthier2007}. 
Of course, extremely large jumps with $\delta_0 \gg \sigma$ would be
unphysical, and would anyway be rejected with high probability 
in dense systems. 

Because the system is now far from equilibrium, equivalence with a
Langevin description is not obvious. Earlier Langevin models 
typically study the following dynamical equations:
\begin{eqnarray}
\frac{\partial}{\partial t} {\vec r_i}(t) & = & v_0 {\vec n_i}(t) + 
\sum_j {\vec f}_{ij}(t) + {\vec \zeta_i}(t), 
\label{eqr} \\
\frac{\partial}{\partial t} \theta_i(t) & = & \zeta^R_i(t),
\label{eqtheta}
\end{eqnarray} 
where $v_0$ is the `bare' 
self-propulsion velocity, ${\vec n_i}=(\cos \theta_i,\sin \theta_i)$
a unit vector whose orientation $\theta_i$ independently 
evolves according to Eq.~(\ref{eqtheta}). 
Note that this set of equations introduces several control parameters, 
having two sources of noise, an energy scale from the interparticle forces, 
and a velocity governing self-propulsion, which typically 
implies that only specific combinations of these are studied. 

It is obvious that, in the dilute limit, such a Langevin description coincides 
with our kinetic Monte-Carlo approach, yielding in particular 
a persistent random walk similar to the data shown 
in Fig.~\ref{fig:MSDPF0}, with a persistent time controlled by the strength of 
the angular noise $\zeta^R_i(t)$ in Eq.~(\ref{eqtheta}).

At finite density, the orientation evolves freely, and the analogy
between Eqs.~(\ref{spp1}) and (\ref{eqtheta}) still holds. However,
the situation is different for Eq.~(\ref{eqr}) 
when interparticle forces compete with self-propulsion and noise.
While displacements in our model are fully slaved to the orientation,
Eq.~(\ref{spp2}), this is not the case in Eq.~(\ref{eqr}) which allows
displacements transverse to the direction imposed by ${\vec n_i}$
resulting from the competition between forces and noise. 
This difference is responsible for the fact 
that some of our results differ from earlier reports. This is
why we introduced the translational noise term $\eta$, as discussed 
above in Sec.~\ref{addingeta}, in order to induce motion in the 
direction transverse to that of the self-propulsion.
We will demonstrate below that this term is crucially needed to trigger a 
macroscopic phase separation.

Recently, 
a noiseless version of the above Langevin equations 
(\ref{eqr}, \ref{eqtheta}) has been studied,    
where ${\vec \zeta_i}= {\vec 0}$~\cite{Filly2012,FilyMarchetti2013}. 
While apparently closer to our 
Monte-Carlo model (the model 
has no translational noise term)
this `athermal' model still allows transverse 
displacements controlled by the force term in Eq.~(\ref{eqr}), at least for 
continuous pair interactions between the particles, showing 
that both random noise and transverse component of the forces 
can induce a macroscopic phase separation. This suggests that the 
origin and statistical properties of the transverse 
component of the displacements are not crucial.

Finally, a model of self-propelled hard particles has recently 
been put forward, which is again reminiscent of our model,
in the sense that it does not introduce any energy scale
through particle interactions~\cite{Ni2013}.
However, in this model a finite amount 
of translational noise is introduced as in Eq.~(\ref{eqr}),
suggesting that this model is in fact closer to our 
generalized Monte-Carlo model where a finite amount of 
translation noise $\eta$ is introduced.

\section{Nonequilibrium structures and phase diagram}

\label{structure}

We start our investigations 
with a detailed structural characterization 
of the stationary phases of the model, exploring in detail 
the phase diagram $(\phi,\tau)$. The effect 
of adding a translational noise $\eta > 0$ is studied 
separately in Sec.~\ref{sec:phsep}.   

\subsection{Phase diagram: Fluid, cluster, and percolated phases}
 
\begin{figure}
\includegraphics[scale=0.47,angle=0]{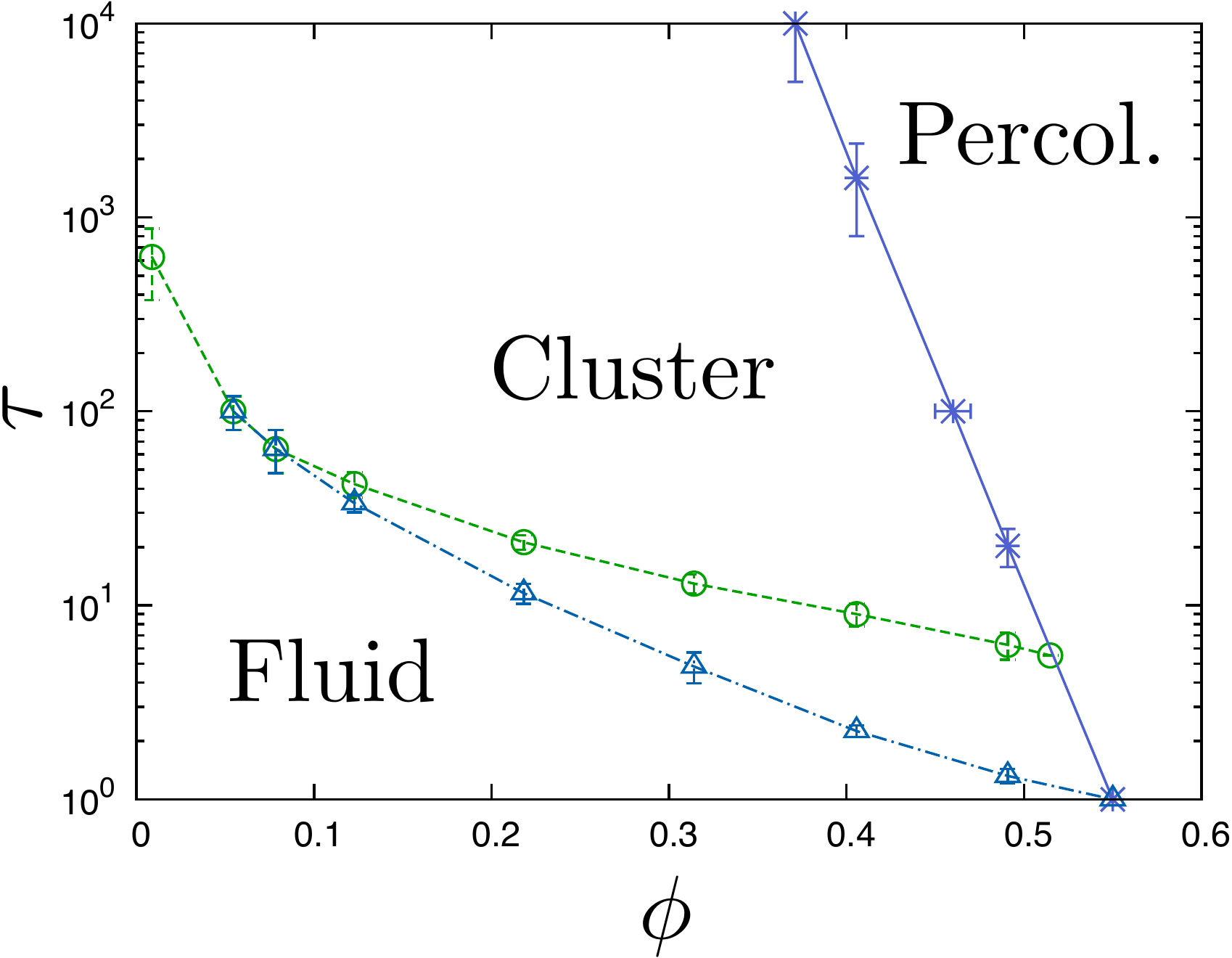}
\caption{Phase diagram of the kinetic Monte-Carlo model of 
self-propelled hard disks in the volume fraction ($\phi$) 
and persistence time ($\tau$) plane. It comprises 
fluid, cluster and percolated phases. The symbols 
are numerical estimates of the phase boundaries.
While the percolation is uniquely defined, the fluid-cluster
boundary is not a phase transition, and we report both a
dynamical (circles) and a structural (triangles) criterion 
to locate this smooth crossover.}
\label{fig:PHD}
\end{figure}

 From a systematic inspection of the steady states
obtained for a large range of values of the 
external parameters $\phi$ and $\tau$,  
we obtained the phase diagram shown in Fig~\ref{fig:PHD}. 
Typical configurations of the system representative of the different phases 
are shown in Fig.~\ref{fig:snapshotsT0}.
In the low-$\phi$, low-$\tau$ region, the system sets into 
a \emph{fluid phase} with the structure and dynamics 
of simple liquids at low density, analogous to a dilute 
suspension of passive disks. This simple phase is expected, as 
we explicitely constructed the model in order to smoothly 
recover the equilibrium situation in the limit $\tau \to 0$. 

Increasing the persistence time at constant density 
induces a strong clustering 
of the particles (see Fig.~\ref{fig:snapshotsT0}) and the system 
enters a phase 
where finite size clusters characterized by a broad distribution of 
sizes coexist in space. We call this a {\it cluster phase}. 
We emphasize that this phase is obtained in a genuine stationary state, 
and that the average cluster size remains finite 
and independent of the time throughout the cluster phase, 
with no tendency towards macroscopic phase separation. 

Clustering in our model occurs naturally as a result of the competition
between self-propulsion (particles move in straight lines)
and hard-core repulsion (particles cannot overlap), so that when 
two particles move persistently towards each other they have to
stop when they touch, until decorrelation of the orientation of the 
displacement vector allows moves that do not create overlaps. Thus, 
as in other models of active Brownian particles, we
expect that self-propulsion is dramatically slowed down in regions 
where the particle density is large~\cite{Tailleur2008}, so that 
active particles have a clear tendency to form aggregates, 
even in systems where interparticle interactions are
purely repulsive.      

The formation of clusters in our model is also 
strongly reminiscent of recent experimental observations 
in systems of self-propelled colloidal 
suspensions~\cite{Theurkauff2012,Palacci2013,Bechinger2013}, 
where motility was shown to produce
finite size clusters, even at moderate densities. 

The two boundaries between the fluid and cluster phases 
shown in the phase diagram in Fig.~\ref{fig:PHD} 
were obtained using two different approaches that we describe 
below. The results show that these two lines 
differ slightly, which underlies the fact that no 
genuine phase transition governs the physics of the cluster 
phase. In other words, the change between fluid and cluster
regions in the phase diagram is a smooth crossover.   

If the cluster phase obtained at moderate density and 
large self-propulsion is now compressed at fixed persistence 
time, the distance between the clusters decreases until a density is 
reached where the clusters percolate throughout the system.
This is the {\it percolated phase} in Fig.~\ref{fig:PHD}.  
As discussed below, the separation line between cluster and 
percolated phases corresponds to a genuine nonequilibrium 
geometrical transition, whose location can be sharply defined 
in the thermodynamic limit. In the percolated region, 
the system has a structure similar to the one found
in physical gels, which can be for instance produced in equilibrium 
conditions in systems of attractive colloidal particles. 

This analogy strengthens further the idea that self-propulsion 
in conjonction with hard-core repulsion induces a kind 
of {\it effective attractive interaction} between active 
particles~\cite{Tailleur2008,Bechinger2013}.   
However, our results differ from most earlier numerical studies 
that have reported very little clustering at low and moderate 
densities, followed by a macroscopic phase separation 
occurring at intermediate densities and finite self-propulsion. 
We will analyse below in Sec.~\ref{sec:phsep} 
the possible reasons explaining these differences,
and have already alluded many times to the key role played by 
the translational noise. 

\subsection{Characterization of activity-driven cluster phase}

\label{subcluster}

In this subsection, we characterize in detail the 
structural properties of the cluster phase resulting from the 
competition between hard core repulsion and self-propulsion. 

The formation of particle aggregates is observed in a 
wide range of physical situations~\cite{MeakinBook}, such as gelation, 
nucleation, coagulation, polymerisation...
Simple models for out-of-equilibrium growth, such as diffusion-limited 
aggregation (DLA)~\cite{Witten1981}, give rise to structures  which can 
be described in the framework of fractal geometry~\cite{PhaseTransVol12}. 
Self-propelled disks differ 
from existing models studied in classic aggregation 
problems because the basic ingredients are specific to our model.  
Here, free particles perform a persistent random walk, which 
is stopped when two or more particles collide, as explained above. 
In the cluster phase, a steady state is obtained because 
free particles diffuse and might aggregate to existing clusters, 
but particles at the surface of these clusters can eventually 
escape, when their (slowly diffusing) 
direction of motion  points towards the exterior
of the clusters. Finite size clusters result from a dynamic 
equilibrium between aggregation and escape of the self-propelled 
disks. Such a dynamic process is also observed experimentally
in systems of self-propelled colloids.

We have characterized the geometric 
properties of the obtained clusters in steady state conditions. 
We define a cluster as a set of disks in contact, that is, 
with interparticle distance smaller than $\delta_0$, the 
elementary jump length. We then define the cluster mass 
distribution  $P(n)$ as the normalized histogram
obtained by measuring the number of clusters containing $n$ particles. 

\begin{figure}
\centering
\includegraphics[scale=0.31,angle=0]{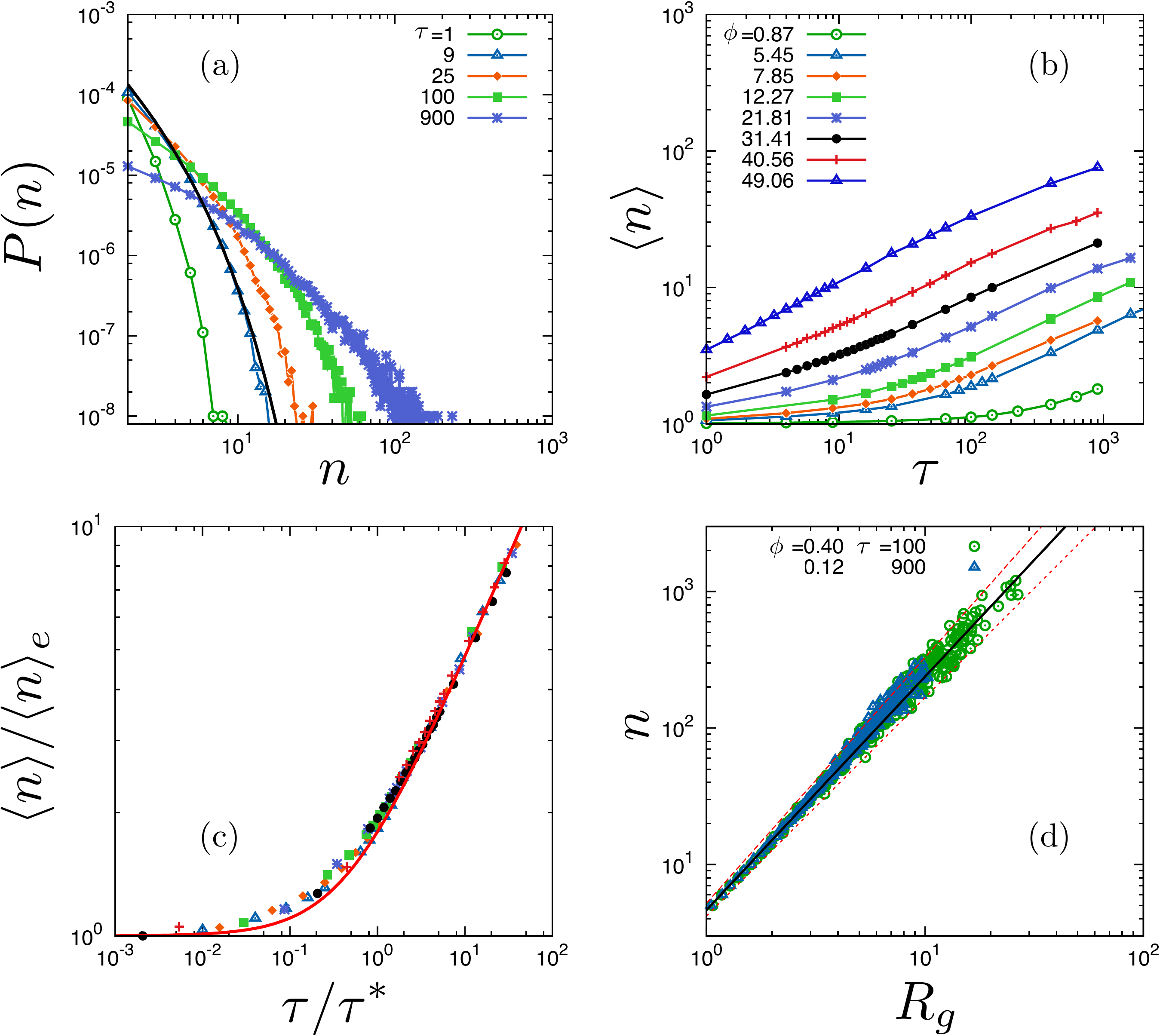}
\caption{
(a) Cluster size distribution for $\phi=0.12$ and different 
persistence times. 
(b) Average cluster size $\langle n \rangle$ as a function 
of the persistence time $\tau$ for different packing fractions 
$\phi$. 
(c) Scaling plot of the average cluster size, rescaled 
by the equilibrium value $\langle n \rangle_e$, while 
the persistence time is rescaled by the crossover value $\tau^*(\phi)$.
The solid red line corresponds to the analytic prediction 
from the kinetic aggregation model, Eq.~(\ref{eq:ms}),  
with $\kappa=3$.
(d) Scatter plot of pairs $(R_g, n)$ for clusters obtained 
at two different state points with measured fractal dimension 
$d_R \approx 1.71$ (full lines). Dashed lines indicate 
$d_R=1.8$ and $1.6$, and provide an estimate for the statistical 
error on the value of $d_R$.} 
\label{fig:cluster}
\end{figure}

In Fig.~\ref{fig:cluster}(a) we show the evolution of 
$P(n)$ obtained for moderate density, $\phi = 0.12$, and increasing the 
persistence time, using a semi-log representation. The effect of increasing 
$\tau$ is clear, as the distribution broadens to include larger clusters.
Regarding the functional form of the distribution, we find that 
it is well described by an exponential form for moderate persistence times, 
but when larger clusters are created, a better functional form is 
\begin{equation}
P(n) \approx n^{-\alpha} \exp \left( - \frac{n}{n^*} \right), 
\end{equation}
where the exponent $\alpha \approx 1.7$, and $n^*$ controls 
the exponential cutoff of the distribution. This functional 
form, shown in Fig.~\ref{fig:cluster}(a) 
has been found in many instances where 
clusters are formed~\cite{Zaccarelli2007RevCollGel,golestanian}. 
It smoothly interpolates between an
exponential form when only small clusters are present, and 
a power law decay, which is found for instance when a percolation 
of the clusters is obtained. At the percolation transition, 
the cluster size distribution becomes scale free, $P(n) \sim n^{-\alpha}$. 
This will be detailed below in Sec.~\ref{percolation}. 

As shown in Fig.~\ref{fig:cluster}(b), the mean cluster mass defined as
\begin{equation}
\langle n \rangle= \sum_{n=1}^{\infty} n 
\,P(n)  ,
\end{equation} 
increases for increasing $\tau$ and $\phi$ in the broad  
range of parameters where steady state properties could be 
explored numerically. 
This is in broad agreement with the images shown in 
Fig.~\ref{fig:snapshotsT0}. 

We can use these measurements to determine a crossover 
line in the phase diagram delimiting the onset of 
clustering. In the fluid phase, we find that
very few clusters are present, such that $\langle n \rangle$ is a slow-varying 
function of $\tau$ and $\langle n \rangle \lesssim 2$.
However, when the persistence time is increased 
the average cluster size starts to increase more rapidly
at a persistence time that depends on the density, because 
clustering becomes more probable at large density. 
A simple argument in the spirit of kinetic theory implies that 
this crossover persistence time should scale as 
$\tau \sim 1/\phi$ when $\phi \ll 1$, 
in good agreement with our numerical data at low density.

The data in Fig.~\ref{fig:cluster}(b) suggest the following 
scaling analysis, that we rationalize below by developing a 
simple kinetic model for aggregation. 
For a given packing fraction $\phi$, we measure the 
average cluster size $\langle n \rangle_e$
for passive disks, i.e. at equilibrium. We then normalize 
$\langle n \rangle$ by $\langle n \rangle_e$, and 
scale the data obtained for different densities 
using the scaled variable $\tau / \tau^*(\phi)$, where 
$\tau^*(\phi)$ is adjusted to produce the best possible collapse. 
The result of this analysis is shown in 
Fig.~\ref{fig:cluster}(c), which demonstrates that 
the scaling form 
\begin{equation}
\langle n \rangle =  \langle n \rangle_e \Phi\left( \frac{\tau}{\tau^*} 
\right)
\label{eqscaling}
\end{equation}
describes the numerical results very well. We find that 
$\Phi(x \ll 1) \sim 1$, whereas $\Phi(x \gg 1) \sim x^{1/2}$
suggesting that the average cluster size increases 
as the square root of the persistence time. 
We rationalize both Eq.~(\ref{eqscaling}) 
and the $\sqrt{\tau}$-dependence in the model described 
below in Sec.~\ref{modelkin}.

An interesting outcome of this procedure is a 
determination, based on structural properties of the system,
of a characteristic persistence time $\tau^* = \tau^*(\phi)$
which marks the onset of clustering. We have reported 
this crossover line using blue triangular symbols in 
the phase diagram shown in Fig.~\ref{fig:PHD}.

To analyse further the (possibly fractal) geometry of the clusters, 
we need to relate the mass of a cluster, $n$, to its size. To this end, 
we measured the radius of gyration $R_g$ of the clusters, defined as 
\begin{equation}
R_g^2(c)=\frac{1}{n_c}\sum_{i\,\in\, c} \left| {\vec r_i} -
{\vec r_c} \right|^2 \, ,
\end{equation}
where the sum runs over all the particles which belong to a cluster $c$ 
of mass $n_c$ and ${\vec r_c}$ denotes the centre of mass
of the cluster.
We also analysed a different length scale, namely the 
end-to-end length $\ell$ of the cluster, which is defined as 
the maximal distance between two particles in the same cluster $c$:
\begin{equation}
\ell (c)= \max_{ \{i,j\} \in \,c} |{\vec r_i}- {\vec r_j} |.
\end{equation}

The dependence of the cluster size, $n$, on these geometrical observables
allows us to explore their  
fractal dimension. We define $d_R$ by $n \sim  R_g^{d_R}$ 
and $d_\ell$ by $n \sim \ell^{d_\ell}$, where $d_R$ and $d_\ell$ are the fractal 
dimensions associated with $R_g$ and $\ell$ respectively. For fractal 
clusters, we expect to find 
$d_R \neq d_\ell<d$ ($d=2$ being the space dimension), meaning that 
the mass grows with a characteristic length  more slowly than 
for compact clusters where $d_R=d_\ell=d$. For instance, for DLA 
in $d=2$ dimensions, $n \sim R_g^{d_{DLA}}$ with 
$d_{DLA}\approx 1.71$~\cite{Witten1981}. 

From the measure of $R_g$ and $\ell$ as a function of the cluster size we 
extract their fractal dimension. We have checked that finite size effects 
do not affect these measurements. As illustrated in 
Fig.~\ref{fig:cluster}(d) for two different state points,  
we obtained the fractal dimensions 
$d_{R}=1.71\pm 0.05$ and $d_\ell=1.60\pm0.05$, independently of 
both the packing fraction $\phi$ and the persistence time 
$\tau$ in the cluster phase. These results strongly suggest that 
clusters in our Monte-Carlo model are always fractal.
Moreover, our simulations indicate that, despite the important differences 
between our model of self-propelled disks and the DLA model,
their fractal dimension $d_R$ is, within our numerical 
accuracy, equal to $d_{DLA}$. As explained above, such an agreement 
is not necessarily expected, as the physics of both aggreation 
processes is not precisely the same, but this does not seem to affect 
profoundly the geometry of the clusters. 

\begin{figure}
\includegraphics[scale=0.47,angle=0]{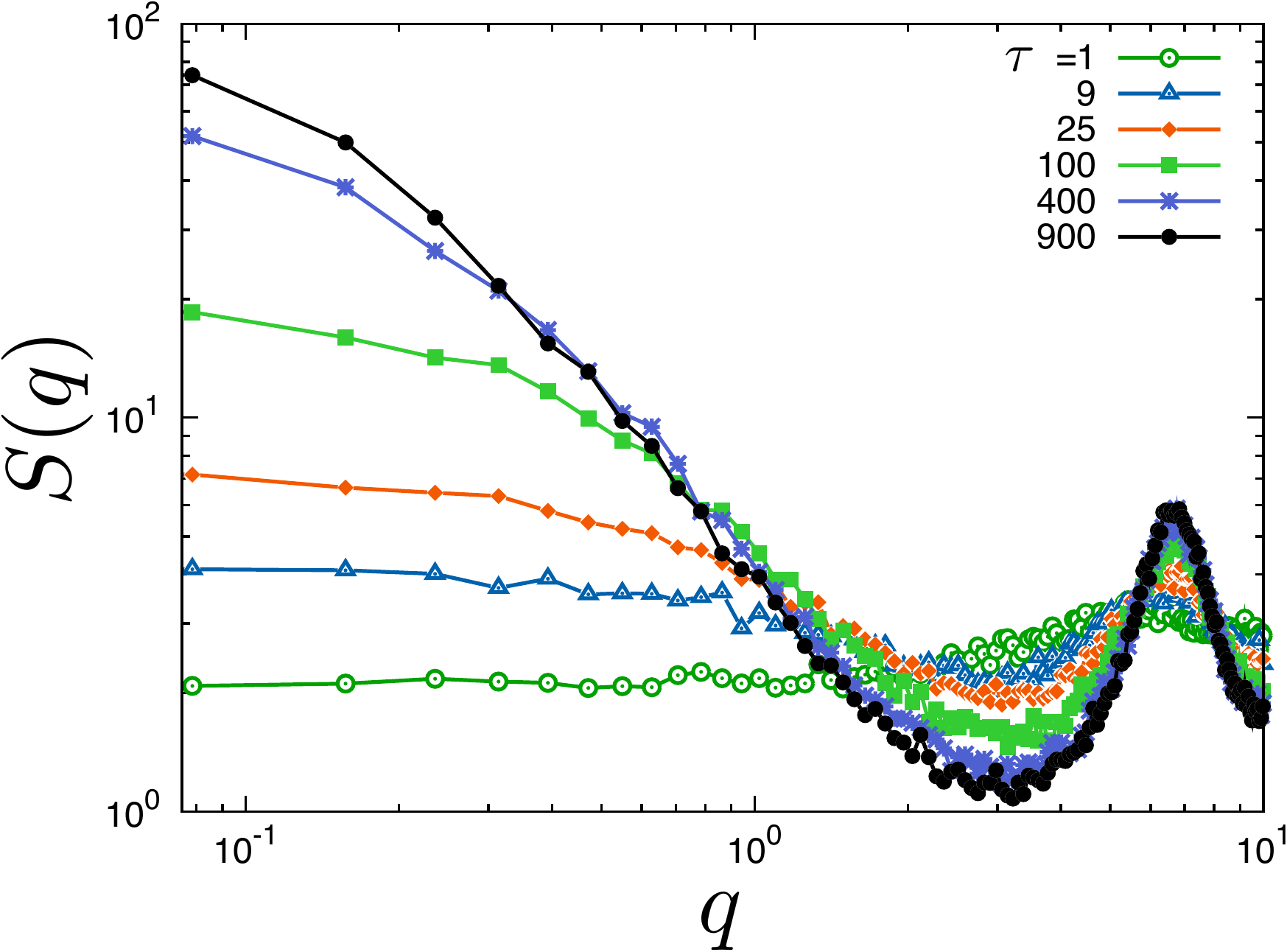}
\caption{Static structure factor, Eq.~(\ref{eqsq}), 
for $\phi=0.12$ and different values of $\tau$. The hard-core 
repulsion alone (small $\tau$) produces a single peak 
at $q \approx 2 \pi / \sigma$, while increasing clustering 
enhances the density fluctuations at much lower wavector, i.e. 
at much larger length scale, revealing the growth 
of the average cluster size.}
 \label{fig:SF}
\end{figure}

Finally, we also characterize the structure of the cluster 
phase using a more standard indicator, namely the static 
structure factor defined as:
\begin{equation}
S(q)=\frac{1}{N} \langle \big|  
\sum_{i=1}^N e^{i {\vec q} \cdot {\vec r_i}} \big|^2 \rangle \, ,
\label{eqsq}
\end{equation}
This observable quantifies the strength of density fluctuations on a length 
scale $\approx 2\pi/|\boldsymbol{q} |$. 
In Fig.~\ref{fig:SF} we show the evolution of the 
static structure factor $S(q)$ for $\phi=0.12$ as the persistence 
time is increased. 
For $\tau=1$, $S(q)$ has the typical 
shape obtained for simple fluids at thermal equilibrium, 
with a maximum at $q \approx 2\pi/r_{n}$, where $r_{n} \sim \sigma$ 
is the typical distance between two neighbouring disks. This peak reflects 
the sole influence of the hard-core repulsion, 
which is the main ingredient controlling the short-ranged 
structure of simple liquids. When increasing the persistence time $\tau$, 
$S(q)$ strongly increases at low-$q$, together with 
peaks characterizing the local structure at wave vectors multiple of 
$2\pi/r_{n}$. This emerging density fluctuations at low-$q$ 
directly reveal the presence of the clusters observed in real space, 
see Fig.~\ref{fig:snapshotsT0}, which produce an inhomogeneous 
structure over a much larger length scale than in simple fluids. 
The increase of the intensity of the peaks at larger $q$ 
reflects the fact that clusters are denser objects than 
the passive fluid. Our structure factors show 
strong similarities with the experimental measurements 
reported in Ref.~\cite{Theurkauff2012} with the  cluster phase  of 
self-propelled colloids.

\subsection{Kinetic model of reversible aggregation}

\label{modelkin}

We now introduce a simple kinetic model to account for the 
evolution of the cluster size distribution with the persistence time, using 
the tools of aggregation models~\cite{RednerKrapivskyBook}.

\begin{figure}
\includegraphics[scale=0.37,angle=0]{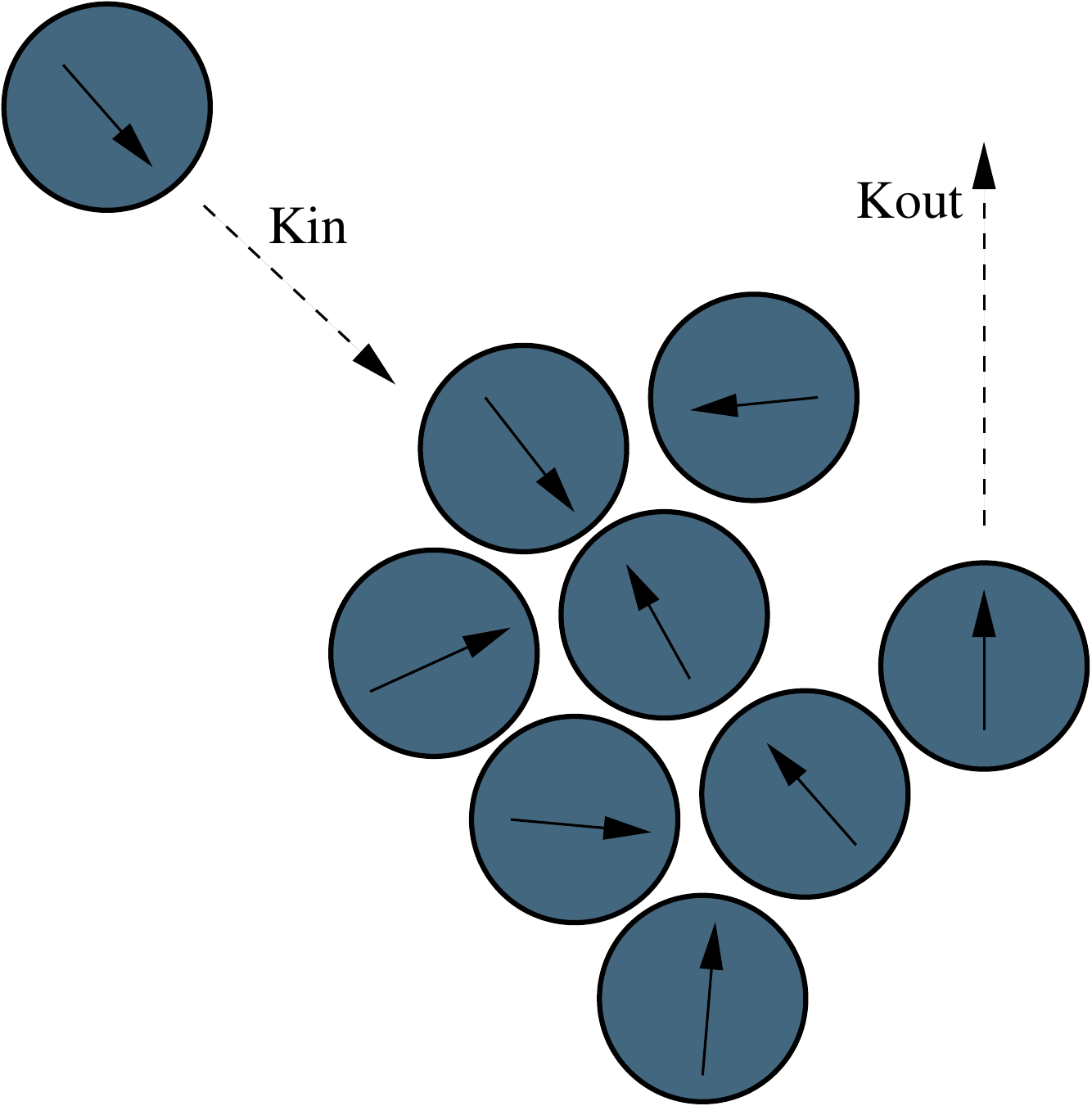}
\caption{Schematic representation of the kinetic model
for reversible aggregation of self-propelled hard disks. 
Particles form clusters because excluded volume 
(represented with blue spheres) opposes self-propulsion 
(indicated by an arrow). Particles outside the cluster can aggregate 
with rate $K_{in}$, while particles at boundaries can escape 
when the direction of motion has diffused and points toward the exterior 
of the cluster, which sets the rate $K_{out} \sim 1/\tau$.}
\label{fig:clustermodel}
\end{figure}

We denote by $c_n(t)$ the number of clusters  of mass $n$ at time $t$.  By 
computing the behaviour of $c_n(t)$ in the limit $t\to \infty$, the model  
should give the main features of the steady-state distribution $P(n) 
= n c_n(t \to \infty)$ obtained in our Monte-Carlo simulations. 
Guided by our numerical observations, we make the following 
assumptions. We assume that the size of a given cluster can only 
evolve by adding or losing individual particles, as depicted in 
Fig.~\ref{fig:clustermodel}. Although seemingly reasonable,  this assumption
implies that we neglect more complicated events such as the aggregation 
of two clusters, or the breaking of a large cluster into two smaller ones. 
Therefore, we consider the 
two following kinetic processes. 
(1) Aggregation of a single self-propelled disk to an existing   
cluster of mass $n \geq 1$ with a rate $K_{in}(n)$. 
The aggregation rate is estimated by assuming that 
particles move ballistically at velocity $v_0$, so that
$K_{in} \propto \phi v_0$. Note that $K_{in}$ does not depend on the 
persistence time, see Fig.~\ref{fig:clustermodel}.
(2) Evaporation of a self-propelled disk from a cluster of 
mass $n>1$ with a rate $K_{out}(n)$. The rate of evaporation is given by 
$K_{out} \propto  1/\tau$: it takes about a persistence time 
$\tau$ to change the orientation of the direction of motion of particles 
in the boundaries of the cluster in order to escape~\cite{Redner2013}, 
see Fig.~\ref{fig:clustermodel}. In the following we also assume a 
mass-independent aggregation and evaporation probability.

In the mean-field approximation neglecting fluctuations, this 
dynamics can be summarized as
\begin{align} 
& [1]+[n] \,{\longrightarrow}\, [n+1] \hspace{0.3cm} \mbox{with rate}\,\,  
K_{in}\, , \\
& [n]\, {\longrightarrow} \, [n-1] + [1] \hspace{0.3cm} \mbox{with rate}\,\,  
K_{out}\, ,
\end{align}
where $[n]$ denotes a cluster of mass $n$.
With these simplifying hypotheses, 
the time evolution of $c_n(t)$ is then ruled by 
a simple set of differential equations:
\begin{align}
\frac{d}{dt} c_1(t)  = & K_{out} \sum_{n=2}^N c_n(t)-K_{in\,}c_1(t)\sum_{n=1}^{N-1} 
c_n(t)\, ,\label{eq:c1}\\
 \frac{d}{dt} c_n(t)  = & K_{in}\left(c_1(t)c_{n-1}(t)-c_1(t)c_n(t)\right) 
\nonumber \\
& +K_{out}\left(c_{n+1}(t)-c_{n}(t)\right)\, .
\end{align}
This is a mean-field model since  we are neglecting the spatial structure 
of the system. The shape  and the location of the clusters cannot be 
described by this approach. A similar modelling has been previously proposed 
to describe clustering in a suspension of elongated self-propelled 
rods~\cite{Peruani2006}. 

In the context of aggregation models, it is common to use an exponential 
\emph{ansatz} for the solutions of the kinetic 
equations~\cite{RednerKrapivskyBook}. Here we consider solutions of the 
following form
\begin{equation}
c_n(t)=(1-f(t))^2f(t)^{n-1} \ ,
\label{ansatz2}
\end{equation}
where the prefactor $(1-f(t))^2$ is found by using the fact that
the number of particles in the system is conserved, $\sum_{n=1}^N 
n c_n(t)=1\, ,\, \forall\ t$. Using the form (\ref{ansatz2}) 
for $c_n$ the kinetic equation (\ref{eq:c1}) becomes:
\begin{equation}
\frac{d}{dt}{f}(t)=(1-f(t))^2-\lambda f(t) \ , \hspace{0.2cm} \mbox{where} 
\hspace{0.3cm} \lambda=\frac{K_{out}}{K_{in}}\, .
\end{equation}
Its general solution reads 
\begin{equation}
f(t)=\frac{1-\exp(-(\alpha-\alpha^{-1})t)}{\alpha-
\alpha^{-1}\exp(-(\alpha-\alpha^{-1})t)}\, , 
\end{equation}
with $\alpha(\lambda) = 
1+\frac{\lambda}{2}+\sqrt{\lambda+\frac{\lambda^2}{4}}$.
The stationary distribution can then easily be obtained by taking the limit 
$t\rightarrow \infty$:
\begin{equation}
\lim_{t\to \infty} f(t)=\alpha^{-1}\, , \hspace{0.4cm} \lim_{t\to \infty}
c_n(t) \equiv c_n^{s}=\lambda \alpha^{-n} \, .
\end{equation}
Note that the stationarity condition $c_1^{s}c_n^s=\lambda c_{n+1}^s$ is 
automatically verified. As expected, the stationary state distribution 
is uniquely 
determined by the relative strength $\lambda = {K_{out}}/{K_{in}}$  
of the two competing processes, whose dependence on the persistence time 
of the self-propulsion is $\lambda = \kappa / \tau$, 
which defines the constant $\kappa$, used below 
as a free parameter.
  
In order to test the validity of this modelling, we compute the mean cluster 
size in the stationary state which is given by the second moment of the 
distribution:
\begin{equation}
\langle n \rangle=\sum_{n>0} n^2 c_n^s=\lambda\alpha^{-1}\frac{1+
\alpha^{-1}}{(1-\alpha^{-1})^3}\, .
\label{eq:ms}
\end{equation}
As shown in Fig.~\ref{fig:cluster}(c), we find an excellent agreement between 
this simple kinetic model and the results of the Monte-Carlo 
simulations of the self-propelled hard disk model, by simply adjusting the 
constant $\kappa = 3$. 
This good agreement 
implies that the dependence of the average cluster size 
is well described by the simple competition between aggregation 
and evaporation of a single particle as sketched 
in Fig.~\ref{fig:clustermodel}, and this predicts 
in particular that the average cluster grows as
$\sqrt{\tau}$ when $\tau \gg 1$. Of course, our mean-field 
model does not provide predictions for the spatial
organisation of the clusters, and cannot in particular
describe the behaviour of the cluster size distribution
when the density increases towards the percolated region. 

\subsection{Percolation transition towards gel-like phase}

\label{percolation}

We mentioned above 
that by compressing the cluster phase above a volume fraction of
about $\phi \approx 0.4$, a percolation transition 
emerges, see the phase diagram in Fig.~\ref{fig:PHD}. This 
transition corresponds to a nonequilibrium version of the contact 
percolation of hard disks~\cite{StaufferBook, Kratky1988, Shen2012},
because configurational sampling is performed far from equilibrium
in the present case. 
Because the cluster size increases with both $\phi$ and $\tau$, 
there comes a point where the clusters touch each other and eventually 
form a system-spanning cluster. This percolation transition 
is reminiscent of the sol-gel transition observed in colloidal
suspensions with attractive interactions. 

\begin{figure}
\includegraphics[scale=0.47,angle=0]{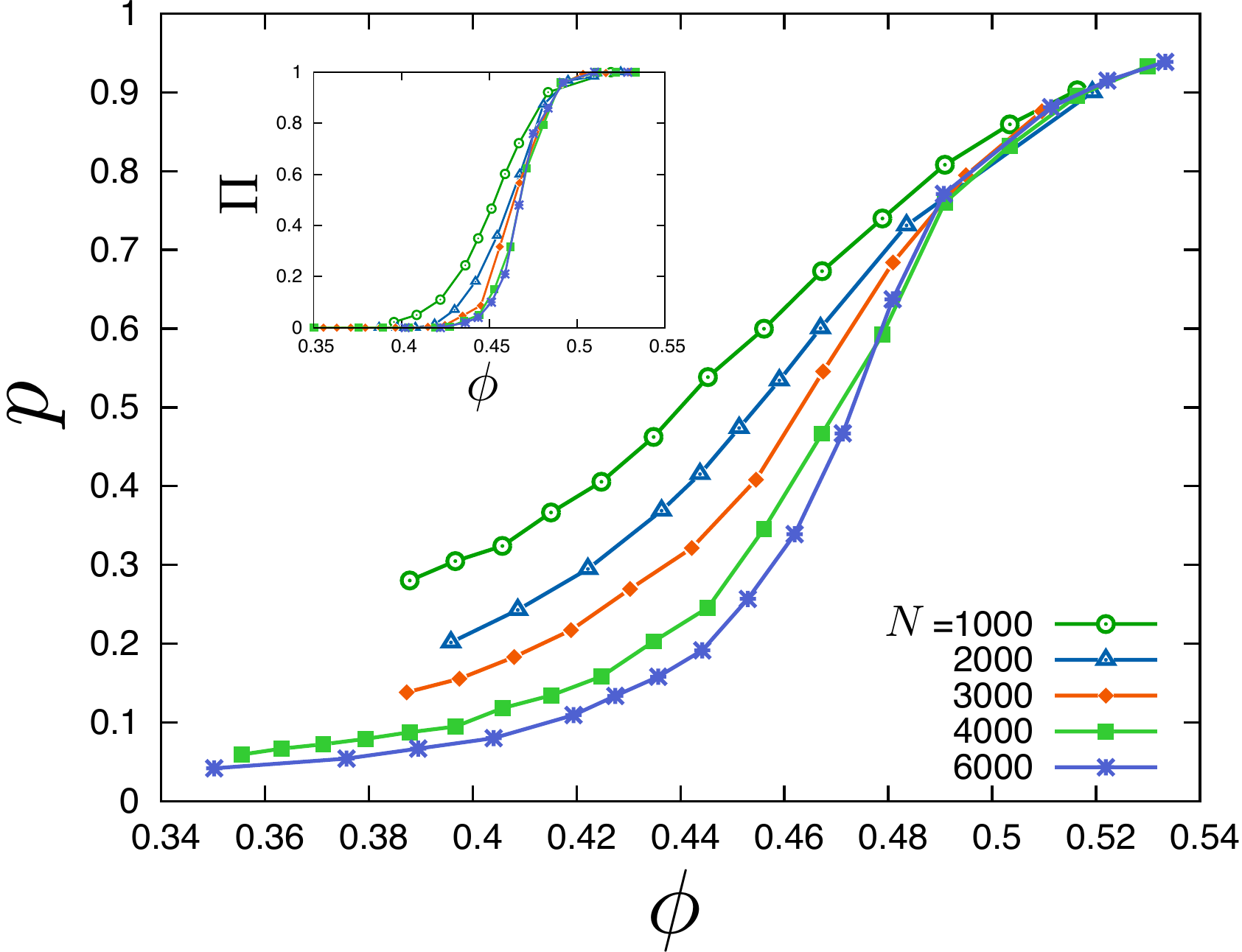}
\caption{Probability $p$ that a particle belongs to the largest 
cluster as a function of $\phi$ for $\tau=100$ and different system sizes. 
The inset shows the probability $\Pi$ that the 
largest clusters does not fit into the simulation box 
as a function of $\phi$ for the same set 
of parameters. Both probabilities exhibit a sharp 
jump from 0 to 1 at the percolation transition in the large $N$ 
limit.}\label{fig:P_R10}
\end{figure}

We use the tools of percolation theory to 
characterise this transition~\cite{StaufferBook,Stauffer1982}. 
In Fig.~\ref{fig:P_R10} we show the 
probability $p$ for a disk chosen at random 
to belong to the largest cluster of the system, measured 
for $\tau=100$ and various system sizes 
$N=1000, \cdots, 6000$. In the inset of Fig.~\ref{fig:P_R10}, we show
the probability $\Pi$ that 
a disk belongs to a cluster which does not fit into the simulation box, i.e. 
such that the end-to-end length $\ell > \sqrt2 L$.  
These two quantities constitute two alternative 
methods to determine the location of the percolation transition, 
as they should jump sharply from 0 to 1 when a percolating cluster emerges
in the thermodynamic limit, $N \to \infty$.  As shown in 
Fig.~\ref{fig:P_R10},  both $p$ and $\Pi$ become sharper 
as system size is increased, although $\Pi$ gives a 
slightly sharper signature 
of the percolation in the vicinity of the 
threshold, which then allows for a more precise determination of 
its location. 
From the data shown in Fig.~\ref{fig:P_R10}, we find a percolation 
transition at $\phi_c(\tau = 100) = 0.46 \pm0.01$. 
In order to construct the phase 
diagram we repeated the same analysis for several values of the parameters 
$\tau$ and $\phi$. These are shown in Fig.~\ref{fig:PHD}. Our data are 
in agreement with the numerical estimation of 
the contact percolation threshold of passive hard disks 
$\phi_c=0.558\pm0.008$ obtained in Ref.~\cite{Shen2012}. 
As expected, we find that the percolation density  
$\phi_c(\tau)$ decreases with increasing $\tau$, which 
mirrors the increase of the cluster size with $\tau$. 
A similar evolution has been reported in reversible 
cluster aggregation models by varying the strength of the 
interaction energy between particles~\cite{Jin1996}, suggesting once more 
an interesting analogy between the role played by the 
persistence time in our model, and the role of attractive forces
in equilibrium systems. 

Another signature of the percolation transition is the 
algebraic decay of the cluster size distribution, discussed 
above in the context of the cluster phase. When percolation 
is approached from the cluster phase, the exponential cutoff 
of the distribution becomes larger, and exactly at 
percolation, we find a purely algebraic cluster 
size distribution $P(n)\sim n^{-\alpha}$ with 
$\alpha=1.70 \pm 0.05$. We note that this 
value of $\alpha$ is far from both the value obtained from random 
bond percolation ($\alpha+1 \approx 2.055$)~\cite{StaufferBook} and continuum 
percolation ($\alpha+1 \approx 2.0$)~\cite{Gawlinski1981}, which 
indicates that our cluster size distribution at percolation 
has a different nature from the one found in 
standard percolation models. We shall see below that 
the effect of adding noise is to 
simultaneously make the clusters more compact and to 
bury the percolation transition inside a phase separated region. 

A physically relevant consequence of the percolation 
transition is the existence of a large area in the 
$(\phi,\tau)$ phase diagram where the structure of the system is very similar
to the one of a `physical gel', characterized by a percolated 
open structure such as the one shown in Fig.~\ref{fig:snapshotsT0}
for $\phi=0.49$ and $\tau=900$. It is interesting to note that
physical gels are obtained in equilibrium colloidal systems with a careful 
tuning of interparticle forces, typically using particles
with both a strong hard core repulsion  and a 
very short-ranged 
attraction~\cite{Weitz2006,Weitz2008,Zaccarelli2007RevCollGel}. 
It is therefore remarkable that 
similar gel-like structures might spontaneously emerge 
using purely hard-sphere interactions in conjunction with self-propulsion. 
 
Interestingly, experiments performed with self-propelled colloidal particles 
have reported the existence of dense phases with large `holes', 
with a structure that is in fact 
strongly reminiscent of our findings, but this phase has not 
been characterized in any detail yet~\cite{isaakThesis}. We hope our 
results will motivate further 
experimental studies of the phase behaviour of self-propelled particles 
in dense regimes. 

\section{Effect of translational noise:  
macroscopic phase separation}

\label{sec:phsep}

We now study the effect of adding a
translational noise of strength $\eta$ in 
our kinetic Monte-Carlo model, as described in Sec.~\ref{addingeta}.

\begin{figure}
\includegraphics[scale=0.5,angle=0]{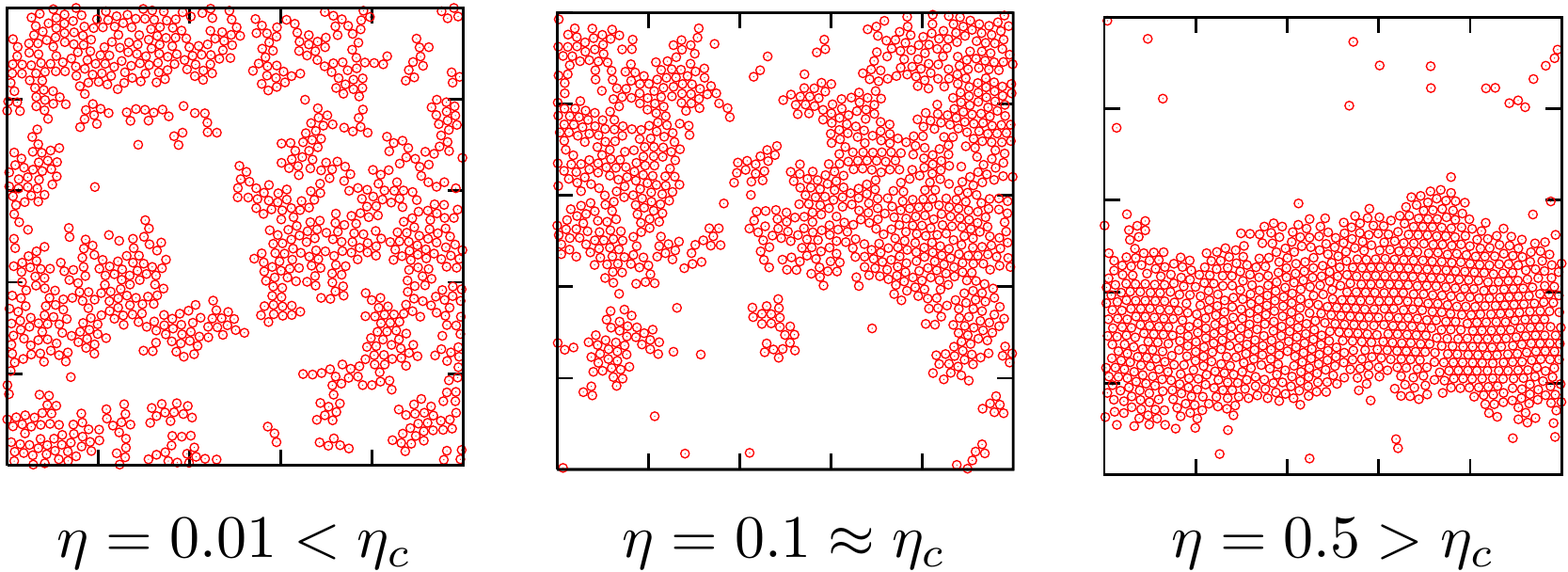}
\caption{Snapshots of the steady state obtained for 
$\phi=0.31$, $\tau=900$ and different noise strength $\eta$.
The translational noise favours the 
formation of more compact clusters, and eventually yields 
a macroscopic phase separation above a critical value
$\eta_c = \eta_c(\phi,\tau)$. }
\label{fig:snapshotsT}
\end{figure}

A first important result is the existence of a finite noise strength
$\eta_c = \eta_c(\phi,\tau) > 0$ below which the phase diagram
obtained for $\eta=0$ remains qualitatively unaffected. This implies 
that our findings are robust against the addition of a finite amount 
of translational noise, see Fig.~\ref{fig:snapshotsT}. 
In particular, this means that adding translational noise 
does not prevent self-propelled disks from aggregating into clusters, 
nor does the noise necessarily induce a macroscopic phase separation,
showing that our results are not `artefact' of our model where
particle displacements are fully slaved to the dynamics of 
the orientation vector.  

However, for $\eta > \eta_c$, we find that the system undergoes 
a macroscopic phase separation reminiscent of liquid-gas demixing
(see the rightmost snapshot in Fig.~\ref{fig:snapshotsT}).
For finite system sizes, we find that the system always reaches 
a steady state comprising a single dense domain surrounded 
by a dilute gas of active disks. Some   
representative snapshots of this situation are shown 
in Fig.~\ref{fig:snapshotsT0} for $\eta = 1/2$. 
We checked that this conclusion is robust if the system size is increased, 
although it takes an increasing simulation time to obtain a single 
domain in larger systems. We call this region of the 
three-dimensional phase diagram $(\phi, \tau, \eta)$ 
the \emph{phase-separated} region (PhS).

It is not straightforward to determine 
quantitatively the transition noise $\eta_c(\phi,\tau)$
by direct visualisation, see Fig.~\ref{fig:snapshotsT}.
Therefore we numerically determine $\eta_c$ by studying
the evolution of the cluster size distribution $P(n)$. 
For $\eta< \eta_c$,  the distribution is dominated by small clusters and 
has an exponentially decaying form. Above $\eta_c$ the size distribution is 
dominated by a macroscopic cluster, so that the shape of $P(n)$ changes 
dramatically from unimodal to bimodal. We define $\eta_c$ by the emergence of 
a second peak in $P(n)$ at large cluster sizes. From a systematic inspection 
of $P(n)$ and direct visualisation of the steady state configurations we 
obtained the value of $\eta_c$ for a broad combination of 
$\phi$ and $\tau$.    

\begin{figure}
\includegraphics[scale=0.35,angle=0]{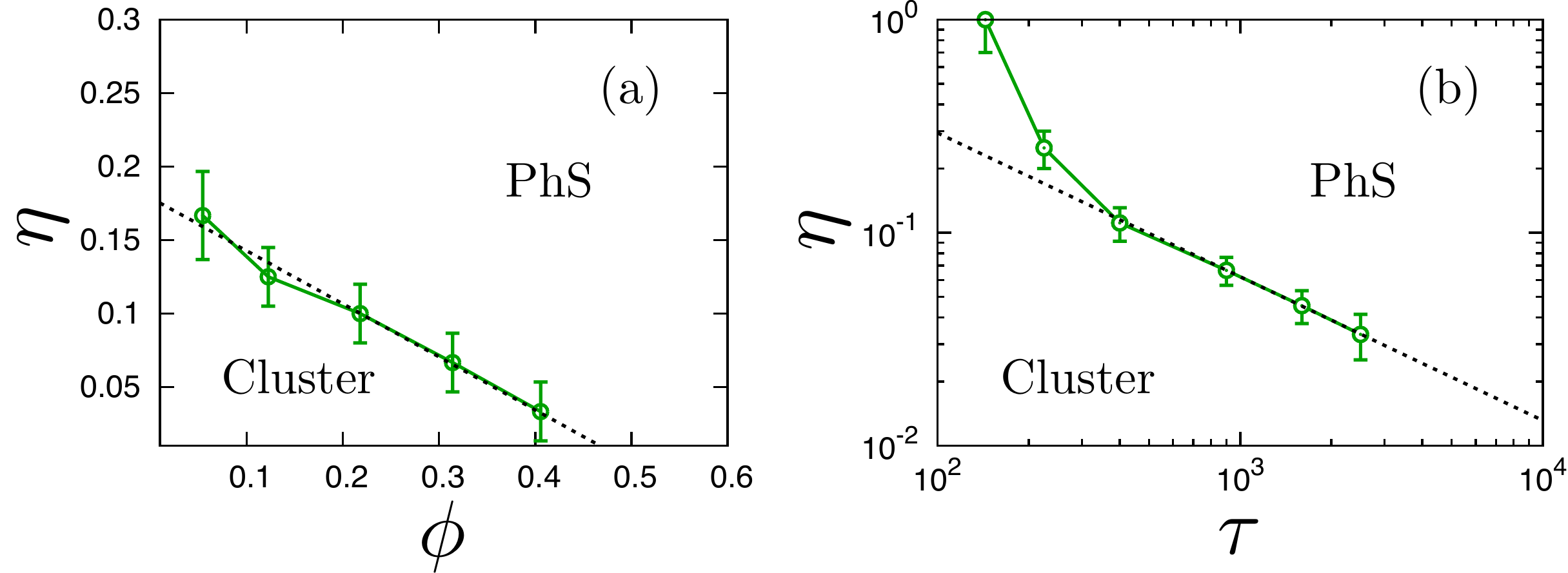}
\caption{Phase boundaries between cluster and phase separated 
phases (PhS) with a finite noise $\eta$. 
(a)  Evolution of $\eta_c$ with volume fraction $\phi$ 
for fixed $\tau=900$ in linear scale showing that 
a macroscopic phase separation can be obtained down to very low 
densities. 
(b) Evolution of $\eta_c$ with persistence time 
$\tau$ for fixed $\phi=0.31$ in log-log scale. No phase separation 
is found for $\tau \leq 100$.  The dotted lines are guides to 
the eye.}
\label{fig:PHD_T}
\end{figure}

In Fig.~\ref{fig:PHD_T}  we show  
representative phase boundaries $\eta_c(\phi,\tau)$ 
between the phase separated region and the cluster phase 
for 
fixed $\tau=900$ over the range $0<\phi<0.4$, Fig.~\ref{fig:PHD_T}(a), and 
for fixed value 
of $\phi=0.31$ over the range $1<\tau< 2500$, Fig.~\ref{fig:PHD_T}(b).
For small persistence times, $\tau \leq 100$, we find that 
macroscopic phase separation never occurs. When $\tau > 100$, we
find that $\eta_c$ decreases when either $\tau$ or $\phi$ 
is increased, see Fig.~\ref{fig:PHD_T}. Moreover, we observe
that $\eta_c$ depends relatively weakly on density, while it is 
changing quite rapidly with the persistence time $\tau$.
Finally, note that when phase separation occurs, it 
exists down to extremely low densities, in contrast with earlier reports
that phase separation only exists at relatively large volume fractions 
in active Brownian 
particles~\cite{Filly2012,Redner2013,Bechinger2013,FilyMarchetti2013}. 

From the snapshots in Fig.~\ref{fig:snapshotsT}, one can expect the geometry 
of the macroscopic cluster in the phase separated region to be different from 
the one in the cluster phase. We have measured the fractal 
dimension of the macroscopic cluster at $\eta > \eta_c$ 
for several systems of different size containing up to $N=6000$ disks. 
We found fractal dimensions $d_R \approx d_\ell \approx 2$ 
in the phase separated region. 
When the system phase separates the resulting macroscopic cluster becomes 
compact, with $n \sim R_g^2$. This is illustrated by the snapshots in 
Figs.~\ref{fig:snapshotsT0} and~\ref{fig:snapshotsT}.
Interestingly, in the region $\eta < \eta_c$, the fractal dimensions 
$d_R$ and $d_\ell$ appear to increase continuously from their 
value at $\eta=0$, $d_R \approx 1.7$ and $d_\ell \approx 1.6$.
 For a given 
$\phi$ and $\tau$, clusters at $\eta=0$ are more ramified than 
at finite $\eta$.
These results are reminiscent of the 
variation of the fractal dimension of the clusters obtained 
by reversible cluster aggregation models with finite bond 
energy~\cite{Jin1996}. Moreover, note that the dense phase at 
$\eta > \eta_c$ appears to be highly ordered, 
contrarily to the fractal clusters obtained at $\eta=0$
that are characterized by a very disordered structure. 

Our central result in this section is that specific combinations  
of self-propulsion, hard-core repulsion, and a finite 
amount of translational noise 
induce a macroscopic demixing as in equilibrium particle 
systems with attractive forces. This conclusion is in broad agreement with 
the phenomenon of motility-induced phase separation 
that has been thoroughly discussed in the 
literature~\cite{Tailleur2008,Thompson2011,Filly2012,CatesRev,Bialke2013EPL,Cates2013EPL,Marenduzzo2013,Stenhammar2014,Wittkowski2013,Redner2013}.
Although the interactions between particles are short-ranged 
and repulsive, the self-propelled disks behave as if there 
were an \emph{effective attraction} between them 
controlled by the persistence time $\tau$. A large amount
of work has also been devoted to the kinetics 
of the phase separation, finding again strong connections
with the equilibrium liquid-gas 
demixing~\cite{Redner2013,Marenduzzo2013,Bialke2013EPL}.

However, our results shed new light on this nonequilibrium
phenomenon. First, we find that translational noise 
is an essential ingredient to trigger the phase separation. 
Physically, our interpretation is that the interplay between 
self-propulsion and hard core repulsion is the primary cause explaining
the emergence of clustering: two hard particles moving in persistent 
motion towards one another cannot cross, and then 
they have to stop (or at least 
slow down) when they approach, which results in enhanced clustering. 

However, our model 
with $\eta=0$ shows that this ingredient is not sufficient
to trigger phase separation. Indeed, we find that fluctuations 
of particle displacements with respect to the direction motion 
imposed by the self-propulsion is a second crucial ingredient,
as it allows {internal relaxation inside the clusters.} A consequence 
of these small intra-cluster motions is that clusters become 
denser and more compact objects. As a result, it becomes 
more difficult for particles to escape the clusters, while 
aggregation remains as easy. This argument suggests that the 
dynamic balance between aggregation and evaporation, which is
responsible for the existence of the cluster phase, 
{\it can be destablized by the addition of noise,} to the point that a 
macroscopic phase separation might occur when aggregation becomes easier than
escape. Therefore, we can view the cluster phase in the original 
model with $\eta = 0$ as resulting from a kinetically arrested phase 
separation.

Another interesting outcome of our study is that 
phase separation might occur over a broad range of densities 
and persistence times, including very low volume fractions.
Alternative models in the literature devised to study 
this phase separation typically do not find a macroscopic 
demixing below $\phi \approx 0.3$, a lower bound 
which is not present in our 
approach~\cite{FilyMarchetti2013}. Overall, these findings 
suggest that macroscopic phase separation is not 
necessarily present in self-propelled particle systems, 
but in fact result from a delicate balance between 
activity, hard-core repulsion and translational noise. 
As explained in Sec.~\ref{comparisonmodel}, previous 
modelling have indeed studied specific combinations 
of these three ingredients, by using for instance specific 
rules for translation/rotation couplings~\cite{Filly2012,Redner2013}. 
However, our results 
show that this balance can be tailored with a greater variety 
than what has been hitherto achieved. 

Because the details of the self-propulsion mechanism
seem to matter even in the context of very simple models, 
it should come as no surprise that specific 
experimental realisations of self-propelled colloidal 
systems find in some cases a macroscopic 
phase separation~\cite{Bechinger2013}, or in some 
other instances the existence of clustered phases without 
a phase separation ~\cite{Theurkauff2012,Palacci2013}. 

\section{Microscopic dynamics  of self-propelled disks}

\label{dynamics}

In this section, we turn our attention to the microscopic dynamics 
in the various phases described above, exploring the 
evolution of dynamic properties as $\phi$ and $\tau$ are varied 
over a broad range, in the original version of the model, i.e. without 
translational noise ($\eta=0$).   

\subsection{Nonmonotonic evolution of the 
diffusion constant}

We have described in Sec.~\ref{subdilute} the microscopic 
dynamics of the system of self-propelled hard disks in the dilute 
limit, $\phi \to 0$. The behaviour is that of a persistent 
random walk with a long-time diffusion constant that increases
with the persistence time as $D \sim \tau$, so that dynamics
becomes faster when $\tau$ increases. 

\begin{figure}
\centering
\includegraphics[scale=0.47,angle=0]{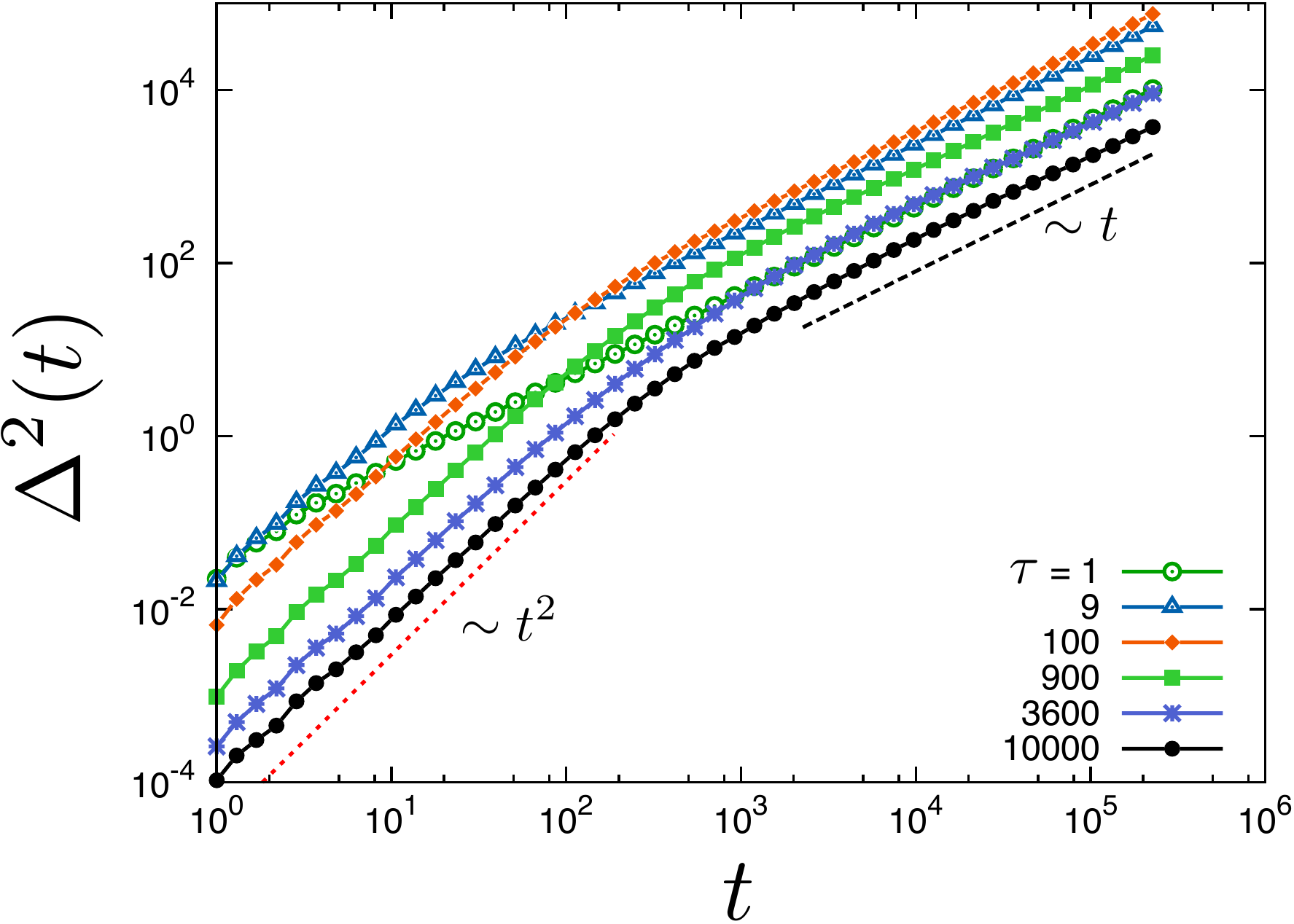}
\caption{Mean-squared displacement, Eq.~(\ref{msd}), 
for $\phi=0.12$ and different persistence times $\tau$. 
Dashed lines indicate ballistic ($\Delta^2 \sim t^2$) and 
diffusive ($\Delta^2 \sim t$) regimes. Note the reentrant 
behaiour of the dynamics that is initially enhanced 
by increasing $\tau$ up to $\tau \approx 100$, 
but slows down when $\tau$ increases further.} 
\label{fig:D_R}
\end{figure}

We now investigate the dynamical behaviour at finite 
density, $\phi > 0$, when interparticle interactions 
play a role. In Fig.~\ref{fig:D_R}, we show the evolution 
of the mean-squared displacement, Eq.~(\ref{msd}), for 
a finite density $\phi = 0.12$, and increasing persistence times.
As expected for the fluid phase, the behaviour 
remains purely diffusive, just as in equilibrium. More interesting
is the behaviour when $\tau$ becomes larger, since we 
find a dynamics similar to that of the dilute limit, with a nearly 
ballistic regime at short-times crossing over to diffusive
behaviour at long-times that is enhanced with respect to the 
equilibrium case.  

The influence of the interparticle interactions is however 
very striking. As shown in Fig.~\ref{fig:D_R}, we find that 
particle displacements are initially enhanced by increasing the 
persistence time of the self-propulsion, as in the dilute limit.
However, a reentrant behaviour is observed when 
increasing $\tau$ further, and long-time dynamics {\it becomes slower}  
with increasing $\tau$. Therefore, we conclude that 
increasing self-propulsion might actually slow the dynamics down, 
which is not a very intuitive result at first sight since it 
does not happen in the dilute system.

\begin{figure}
\centering
\includegraphics[scale=0.48,angle=0]{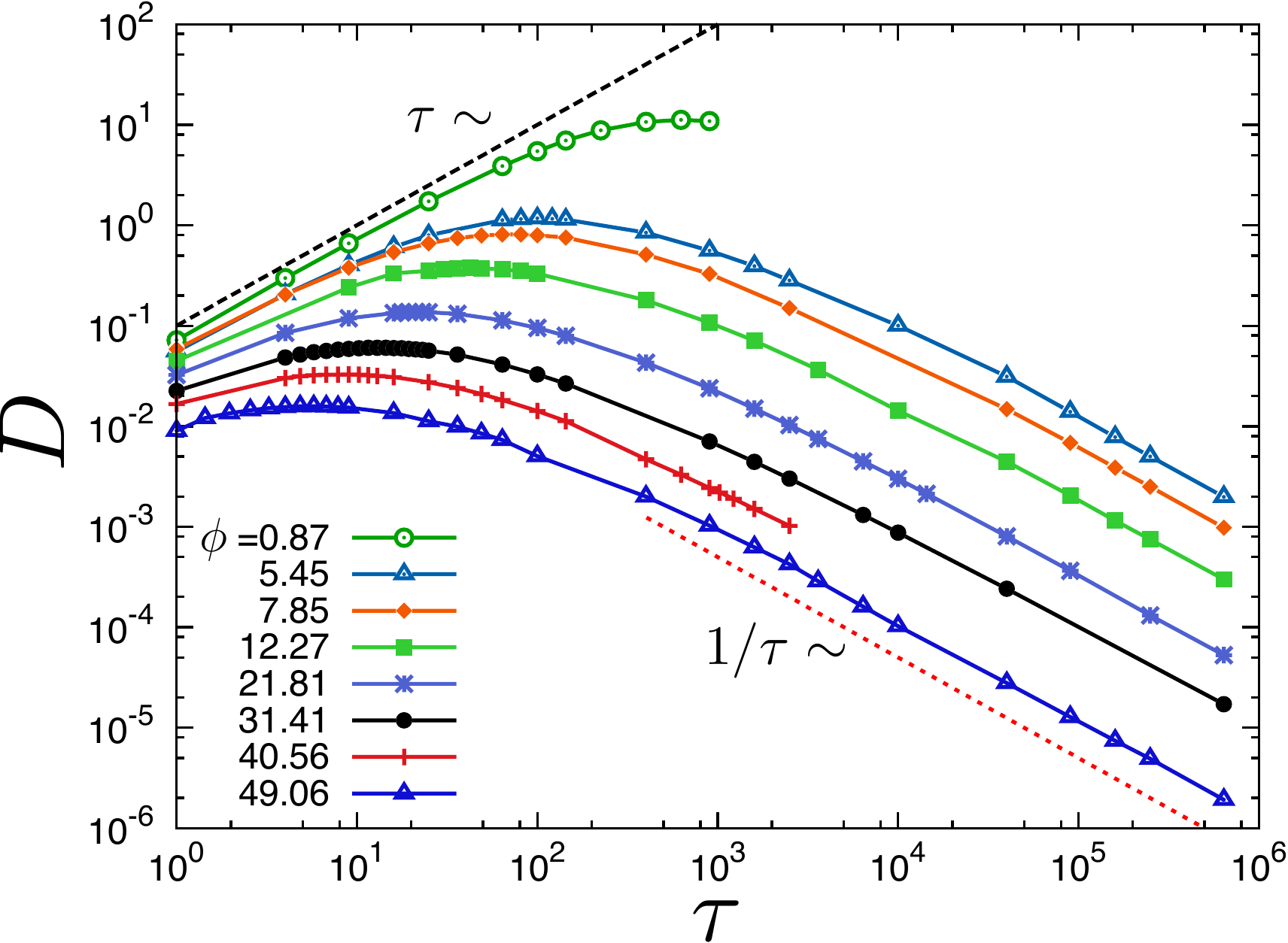}
\caption{Diffusion 
coefficient as a function of the persistence time 
$\tau$ for different values of the packing fraction $\phi$. 
The dashed lines describe the behaviour in the dilute limit 
$D \sim \tau$ which holds when $\phi$ and $\tau$ are small, 
while the $D \sim 1 / \tau$ describes the large-$\tau$ limit 
at any finite density, as a consequence of clustering. 
The crossover between these two regimes
defines an optimal persistence time $\tau_d(\phi)$ where 
diffusivity is maximised.} 
\label{fig:D_R2}
\end{figure}

To quantify this effect further, we measure the diffusion 
constant defined as 
\begin{equation}
D(\phi,\tau) = \lim_{t \to \infty}\frac{\Delta^2(t)}{4 t}\, .
\end{equation} 
We present results for the evolution of the diffusion constant 
with the persistence time at various volume fractions in 
Fig.~\ref{fig:D_R2}.  The dilute 
limit behaviour $D \sim \tau$ is recovered for small packing fractions and 
persistence times. However, for any finite density the diffusion 
constant exhibits a maximum, and it decreases at large 
$\tau$, with the asymptotic behaviour given by 
$D \sim 1/\tau$ as $\tau \to \infty$. 
The maximum of the diffusion constant suggests that there
exists an `optimal' value of the persistence time $\tau_d = \tau_d(\phi)$ 
which maximises the diffusivity for a given 
packing fraction $\phi$. The data in Fig.~\ref{fig:D_R2}
suggest that $\tau_d$ decreases with increasing the density.   

The interpretation of the maximum of diffusivity is 
quite straightforward, because it is a direct consequence 
of the emergence of the clustered structures described in 
Sec.~\ref{structure}.  At small $\tau$ and $\phi$, when small 
clusters start to form, the spatial structure becomes heterogeneous.
This opens large voids where self-propelled particles can move almost 
freely, i.e. as in the dilute limit. Accordingly, 
the diffusion constant increases
with the persistence time in this regime. 
However, when $\tau$ increases further, the clusters may become large.
The key point is that particles deeply buried inside 
the clusters can be trapped there for long 
periods of time.  In this regime, increasing $\tau$ 
can have two opposite effects. Particles 
in the dilute phase move faster, but particles inside the clusters 
are arrested for increasing periods of time. Our simulations indicate
that when $\tau$ becomes very large, most particles 
are arrested inside clusters, and the second effect dominates,
which explains why $D$ eventually 
decreases with increasing the self-propulsion. A similar
effect has been recently reported for the mobility of 
active particles driven through a disordered medium~\cite{reichhardt2014active}.

This optimal persistence time $\tau_d$ 
corresponds therefore to a delicate balance between 
two competing effects: increasing $\tau$ accelerates the dynamics
of individual particles, but also produces clustered 
structures where particles are kinetically trapped. Because clusters 
form more easily at higher packing fractions, the optimal persistence value 
$\tau_d(\phi)$ decreases when $\phi$ increases.

The optimal value $\tau_d$ offers an alternative definition 
for the location of the crossover between fluid and cluster phases, 
as it provides a dynamical signature of the emergence of clusters.
We report the measured values of $\tau_d(\phi)$ 
obtained from the maximal diffusivity for different packing fractions 
in the phase diagram in Fig.~\ref{fig:PHD}, which gives 
a crossover line that is in good agreement with the one
obtained from directly studying the structure, and already discussed 
in Sec.~\ref{subcluster}.   
 
The last piece of information we need to discuss in 
Fig.~\ref{fig:D_R2} is the observed large-$\tau$ 
behaviour of the diffusion constant, namely 
$D(\tau \gg \tau_d ) \sim 1/\tau$. 
In this regime, particles alternate between few 
periods of fast ballistic motion and long periods of kinetic 
trapping within clusters. 
As usual, the diffusive dynamics is dominated by the fastest 
particles, which are the ones sitting on the surface of 
the clusters, which can escape with a rate proportional
to $1/\tau$, as in the sketch of Fig.~\ref{fig:clustermodel}.
This limiting rate for cluster escape directly 
accounts for the scaling of the diffusion constant, $D \sim 1/\tau$. 

\subsection{Non-Fickian diffusion and decoupling} 
 
\begin{figure}
\centering
\includegraphics[scale=0.47,angle=0]{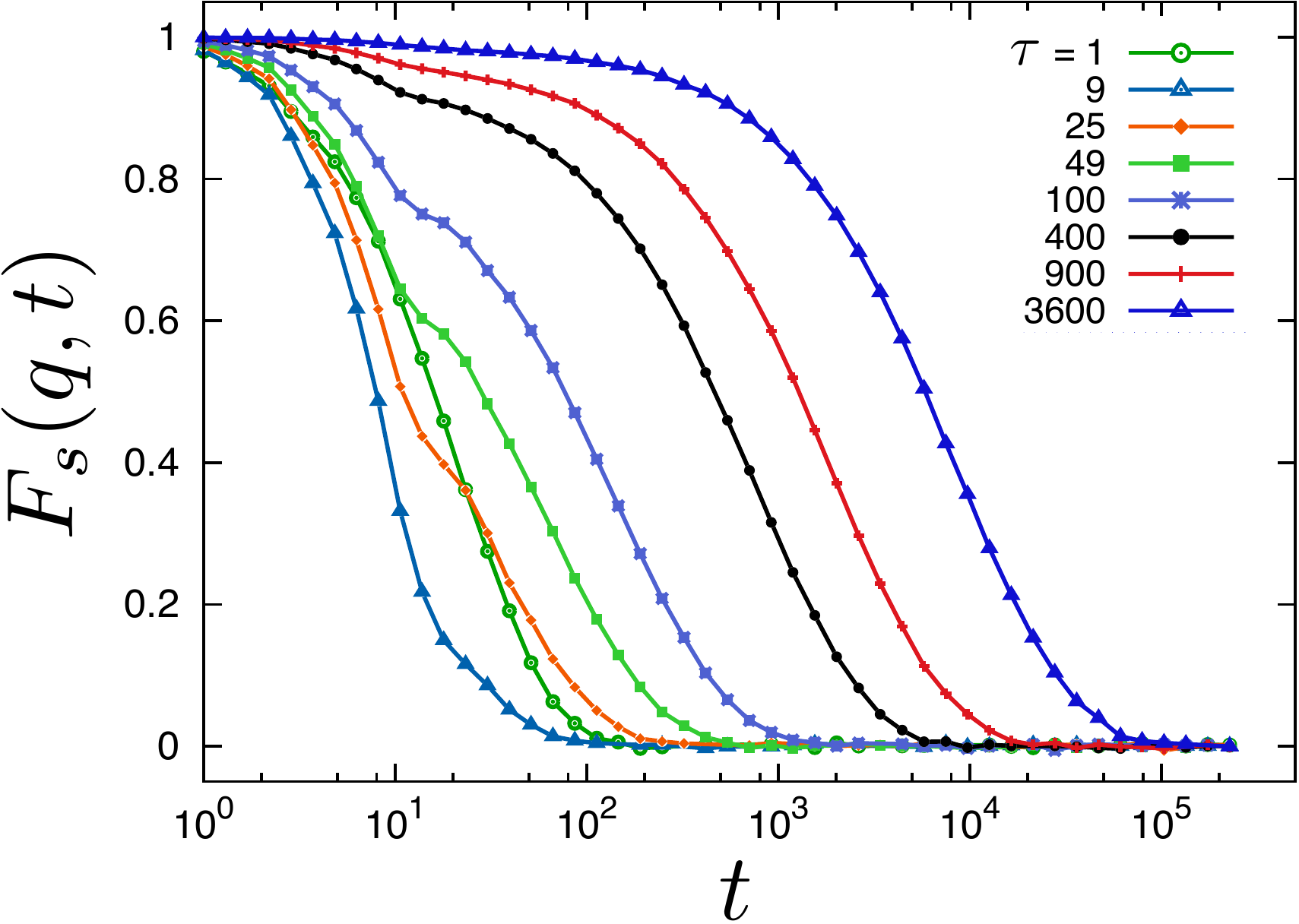}
\includegraphics[scale=0.47,angle=0]{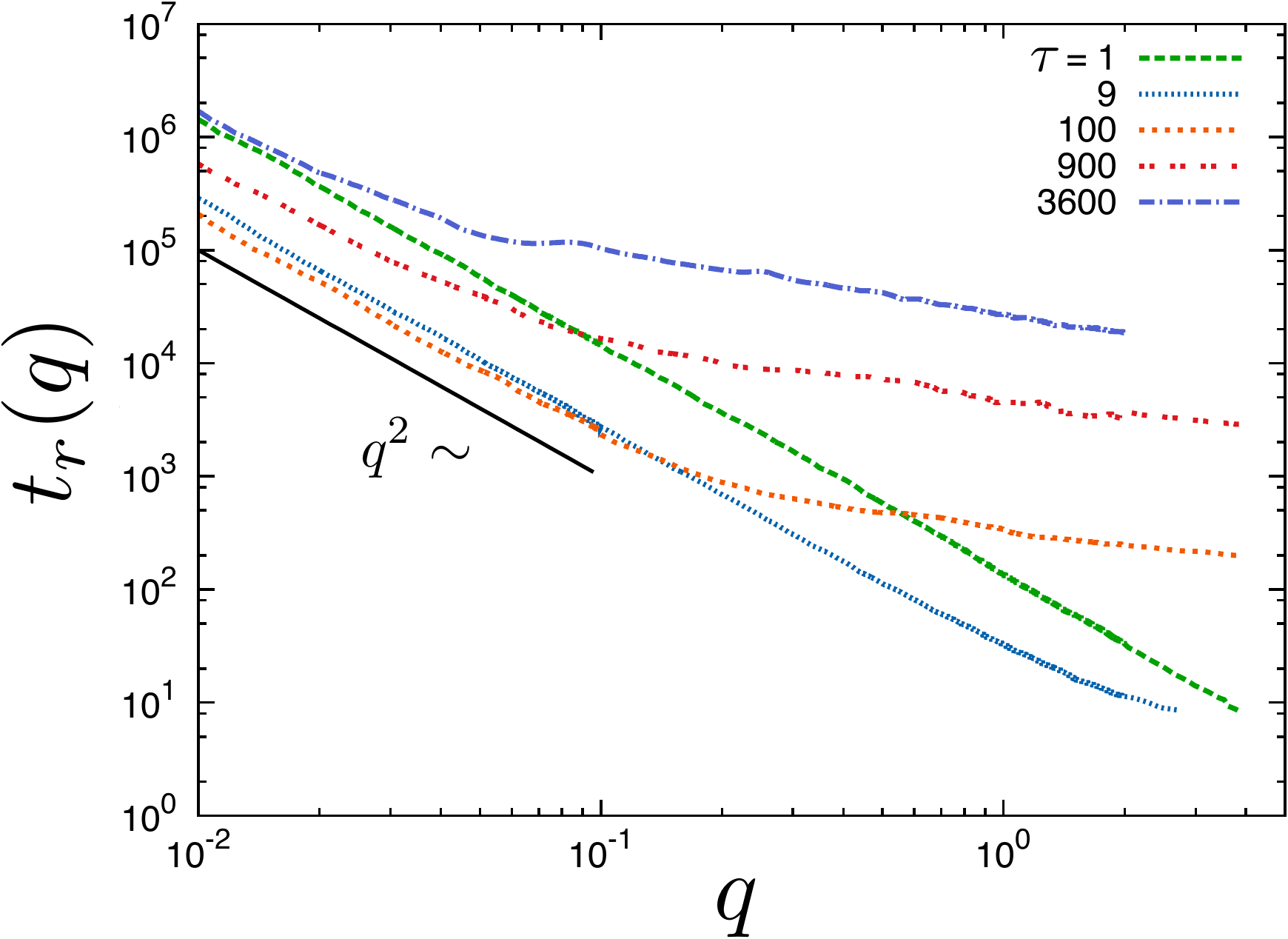}
\caption{Top: Self-intermediate scattering function,
Eq.~(\ref{SIC}) for $q=1.20$ for $\phi=0.12$ and 
different values of the persistence time.
Bottom: Relaxation time $t_r(q)$ for the same density, with 
Fickian behaviour $t_r \sim 1/q^2$ indicated with a dashed line.}
\label{fig:SIC_varR}
\end{figure}

To further characterise the microscopic dynamics, 
we compute additional time correlation functions.
A natural quantity, which is particularly relevant for 
scattering experiments, is the self-intermediate scattering function,
\begin{equation}
F_s( {q},t)=\frac{1}{N}\langle   \sum_{i=1}^N  e^{i {\vec q}
\cdot [ {\vec r_i}(t)- {\vec r_i}(0)] } \rangle \, ,
\label{SIC}
\end{equation}
which quantifies particle motion over a typical length scale 
$\approx 2 \pi / q$.
In Fig.~\ref{fig:SIC_varR}, we show the evolution of $F_s(q,t)$ from 
the fluid phase to the cluster phase at fixed packing fraction $\phi=0.12$. As 
expected for the fluid, 
the relaxation is fast and exponential in the near-equilibrium case, 
$\tau=1$. Increasing $\tau$, we observe a first relaxation towards 
a plateau which emerges at intermediate times (which is not very pronounced), 
followed by a second 
slower relaxation. The height of this plateau strongly depends 
on the value of $\tau$, while its duration does not (and remains 
relatively short). Increasing $\tau$ further, the dynamics 
slows down very rapidly, but the time dependence of $F_s(q,t)$
does not evolve qualitatively. 
This behaviour is strongly reminiscent of the viscoelastic 
relaxation observed in reversible physical 
gels~\cite{Zaccarelli2007RevCollGel,Hurtado2007}. 

Here, the intermediate time plateau emerges as a result 
of the formation of a heterogeneous structure
due to the aggregation of particles. Because both clusters 
(percolating or not) and a dilute phase coexist, there naturally
exist two distinct dynamical families of particles relaxing on 
two different time scales, and the plateau height reflects their 
relative weight. Thus, our dynamical study 
adds one more element to support the physical idea 
that nonequilibrium persistent motion of repulsive particles produces 
a physical behaviour reminiscent of equilibrium particles with 
attractive forces. 

The long-time decay of $F_s(q,t)$ corresponds to the structural relaxation of 
the system. From this decay, one can extract a relaxation 
time scale $t_r(q)$, which we define as  
$F_s(q,t_r(q))=0.2$ (the precise value 0.2 is irrelevant).
It is interesting to study the wave vector dependence 
of the relaxation time,  as displayed in Fig.~\ref{fig:SIC_varR}. 
Whereas purely diffusive behaviour, $t_r \sim q^{-2}$, is observed 
for near-equilibrium dynamics at small $\tau$, the 
structural relaxation time displays a crossover between 
two different regimes when $\tau$ increases. Diffusive 
behaviour is still observed but only when $q$ is very low, whereas the 
behaviour at larger $q$ is very strongly non-Fickian with 
$t_r(q)$ being only weakly dependent on wave vector. 

Deviations from Fickian behaviour are generically 
expected in systems exhibiting strong dynamic 
heterogeneity~\cite{BerthierBook}. In particular, 
systems characterized by the existence of distinct 
dynamic populations (e.g. fast and slow particles) 
with kinetic exchanges between the two populations~\cite{Hurtado2007,pinaki2} 
all exhibit two-step decay in intermediate scattering functions 
and non-Fickian dynamics at short scale of the type shown 
in Fig.~\ref{fig:SIC_varR}. 

The $q$-dependence of the relaxation
is easily explained in this dynamically heterogeneous scenario. 
In a system with fast and slow particles, it is the slow population
which controls the long-time decay of the self-intermediate 
scattering function at large $q$, which basically quantifies 
how long it takes for initially trapped particles to start moving.
On the other hand, the low-$q$ behaviour is controlled
by the long-time diffusion constant, $D$, which does not
necessarily quantify the same dynamic process.
In the two-population scenario, for instance,
$D$ is essentially controlled by the exchange rate between the two families, 
whereas $t_r(q)$ at large $q$ is controlled by the relaxation time
of the slow one~\cite{berthier2005length}. 
These two distinct measures of the 
relaxation time only become equivalent when dynamics 
is homogeneous and purely Fickian, in which 
case one has $t_r(q) \sim 1/(D q^2)$. Our results suggest 
instead that $t_r$ increases much faster than $1/D$ with 
increasing $\tau$. The data in Fig.~\ref{fig:SIC_varR} suggest 
for instance a growth of about 3 orders of magnitude of 
the product $D \times t_r(q)$ 
(for $\phi=0.12$ and $q=1.2$) and $\tau$ increasing by about 
4 orders of magnitude.  
The independent evolution of $t_r(q)$
and $1/D$ is called a `decoupling' phenomenon in the context 
of dynamically heterogeneous materials~\cite{BerthierBook,Ediger2000}. 
 
Non-Fickian diffusion and decoupling phenomena have been 
reported in a large number of physical systems~\cite{BerthierBook,Ediger2000}, 
from supercooled liquids approaching a glass transition
to dense granular, colloidal suspensions, and colloidal 
gels~\cite{Guo2005,Hurtado2007}. For the latter type of systems,
dynamic heterogeneity is a direct (and physically 
unsurprising) consequence of a heterogeneous structure~\cite{gel2,Hurtado2007}, 
as is the case 
for the present model where the structure in cluster and percolated 
phases is directly responsible for the peculiar dynamic 
features reported in this section.  

\subsection{Dynamic heterogeneity}

The decoupling of the evolution of the diffusion 
constant and the relaxation time together with the strongly non-Fickian 
wave vector dependence discussed above suggest that microscopic 
dynamics is spatially heterogeneous in our model of self-propelled 
hard disks. 

A simple way to observe this dynamic heterogeneity more directly
is to focus on distributions of particle displacements. 
We have measured the self-part of the van Hove function, defined as 
\begin{equation}
G_s(x,t)=\frac{1}{N} \langle  \sum_{i=1}^N \delta( x - 
| {x_i}(t) - {x_i}(0) |  )\rangle \, .
\label{Gs}
\end{equation}
Because the system is isotropic, we can average the above function 
along both space directions. 
This function is akin to a (normalized) histogram of single  
particle displacements measured over a fixed time delay, 
which takes a simple Gaussian form for purely Fickian 
dynamics. Note that $G_s(x,t)$ is also related to $F_s(q,t)$ by a 
Fourier transform.

\begin{figure}
\centering
\includegraphics[scale=0.27,angle=0]{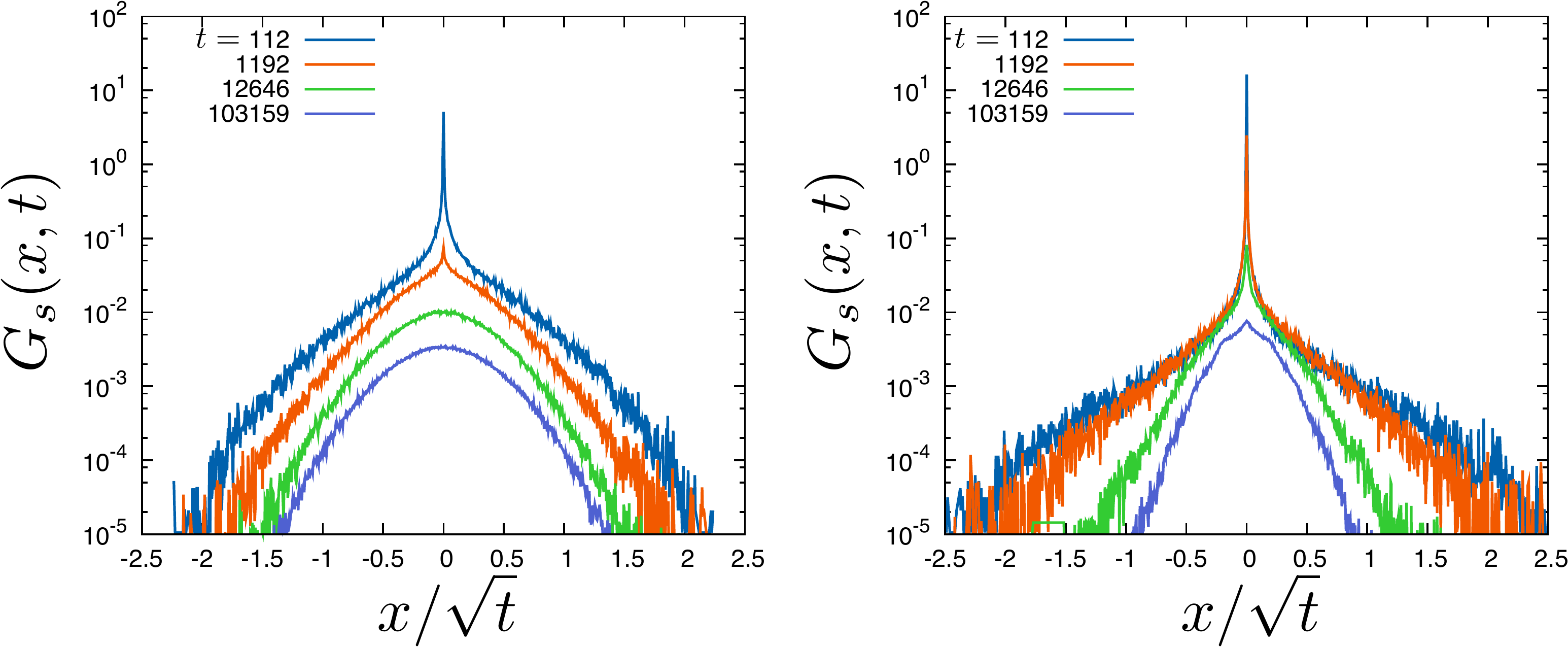}
\caption{Distribution of particle displacements, Eq.~(\ref{Gs}), 
for $\phi=0.12$, and $\tau=100$ (left) and $\tau = 900$ (right).
The distributions are composed of a superposition of immobile
(near $x \approx 0$) and very mobile (in the nearly exponential 
tails) particles, and converge to Gaussian form at long times.}  
\label{fig:Gs}
\end{figure}

In Fig.~\ref{fig:Gs}, we show the time evolution of $G_s(x,t)$ 
for $\phi=0.12$ and two values of the persistence time, 
$\tau = 100$ and $\tau = 900$. To better appreciate 
the time evolution of the shape of these distributions, 
it is convenient to use a rescaled unit $x/\sqrt{t}$, 
as dictated by the long-time diffusive limit. While the distributions
eventually converge to the expected  Gaussian distribution, 
quite strong deviations from Gaussian behaviour are observed 
at finite times. In particular, the coexistence of nearly arrested 
and fast-moving particles is obvious at short times, since 
the van Hove function displays both a very narrow peak at the origin 
stemming from immobile particles, and broad tails (i.e. broader than the 
corresponding Gaussian distribution) stemming from fast particles. 
The distinction between mobile and immobile particles 
becomes more pronounced as the persistent time is increased. 
Such broad tails are observed 
in a large number of dynamically heterogeneous materials~\cite{BerthierBook},
where they are usually found to be described by an exponential 
decay~\cite{chaudhuri2007universal}, 
as is also found in the data shown in Fig.~\ref{fig:Gs},
especially at large $\tau$.  

As time increases, the height of the central peak decreases 
because the population of immobile particles 
naturally decreases with time (eventually at long times 
all particles become mobile). As expected, this arrested 
structure stays immobile over a longer period when $\tau$ increases. 
When all particles have finally moved, the tails of the distribution
become closer and closer to the corresponding Gaussian distribution, 
and dynamic heterogeneity is washed out.  

These distributions confirm that the microscopic dynamics is heterogeneous 
due to the coexistence of arrested particles inside clustered regions, and 
free particles diffusing rapidly in dilute regions. This  
accounts for both non-Fickian diffusion and decoupling 
of diffusion constant and structural relaxation time. 
This physical behaviour is reminiscent of the
dynamic heterogeneity discussed in the context of colloidal 
gels~\cite{gel1,gel2,Hurtado2007}. 
It seems clear that the simple kinetic models developed in the 
context of colloidal gels to describe analytically the shape of 
the van Hove distributions~\cite{Hurtado2007,pinaki2}  
would be directly applicable to the present model. We suggest 
that such models would also be useful in the context of 
experimental investigations of dynamic heterogeneity 
in self-propelled particle systems, but this line of 
investigation has been little explored. 

\section{Summary and conclusion}

\label{discussion}

To summarize, we have introduced and studied in great detail 
a kinetic Monte-Carlo model for self-propelled hard disks, 
which contains a minimal amount of free parameters. 
As a result, the model is quite crude, but it still captures 
essential features of the competition between thermal fluctuations, 
self-propulsion and hard core repulsion between the disks. 
Although seemingly simpler than alternative models in the literature, we
have demonstrated that it is also somewhat more flexible, in the sense
that it allows us to disentangle more clearly the specific roles
played by each ingredient in the model, but also to study 
combinations of parameters that have not been explored 
in earlier models. 

Our main findings are the emergence of non-trivial nonequilibrium 
structures at finite density, which take the form of a cluster phase 
at moderate density, eventually percolating into a gel-like
phase at larger density. Adding a controlled amount of 
translational noise to the model allows us to connect our results 
to earlier studies of a motility-induced macroscopic phase separation. 

We have compared our results to experimental model systems 
of self-propelled colloidal suspensions for which 
both a cluster phase and a phase separated state  have been observed 
as well~\cite{Theurkauff2012,Bechinger2013,Palacci2013}. We note that the 
gel-like phase we obtain at large density has not been studied 
experimentally in any 
detail yet. Our investigations of the microscopic dynamics 
of this phase suggest that it is potentially a very interesting regime, 
characterized in particular by strong dynamic heterogeneities, 
similar to the ones found in colloidal gels formed with attractive 
particles. 
 
In future work, we aim at investigating more closely the 
striking analogies underlined in this work between the nonequilibrium
physics of self-propelled disks and the equilibrium behaviour 
of systems with short-range attractive forces. Additionally, we 
believe the present kinetic Monte-Carlo model is particularly well-suited
to study the effect of self-propulsion at large densities, 
since it captures the essential elements of the competition 
between slow dynamics emerging because of crowding at large density,
and the activity of the particles that provides the fuel needed 
to move them faster. While few 
numerical~\cite{Ni2013,Berthier2013,FilyMarchetti2013} 
and theoretical~\cite{BerthierKurchan,Brader2014MCT} studies of 
this competition have appeared, we also encourage experimental 
work in that direction. 
 
\acknowledgments 
We thank T. Kawasaki for useful exchanges.
The research leading to these results has received funding
from the European Research Council under the European Union's Seventh
Framework Programme (FP7/2007-2013) / ERC Grant agreement No 306845.

\bibliographystyle{apsrev}
\bibliography{AHDshort}

\end{document}